\documentclass[graybox]{svmult}

\usepackage{mathptmx}       
\usepackage{graphicx}                                   
\usepackage{multicol}        
\usepackage[bottom]{footmisc}
\usepackage{bm}
\usepackage{calrsfs} 
\usepackage{amsmath}

\makeindex        

\begin{document}

\title*{What can spin glass theory and analogies tell us about ferroic glasses?}
 \titlerunning{Spin glass theory and   ferroic glasses}
\author{David Sherrington}
\institute{David Sherrington \at Rudolf Peierls Centre for Theoretical Physics, Clarendon Laboratory, Parks Rd., Oxford OX1 3PU, UK 
\at  Santa Fe Institute, 1399 Hyde Park Rd, Santa Fe, NM 87501, USA
\newline
 \email{David.Sherrington@physics.ox.ac.uk}}
\maketitle
\vspace{-75 pt}

\abstract{
As well as several different kinds of periodically ordered ferroic phases, there are now recognized several different examples of ferroic glassiness, although not always described as such and in material fields of study that have mostly been developed separately. In this chapter an attempt is made to indicate common  conceptual origins and features, observed or anticipated. Throughout, this aim is pursued through the use of simple models, in an attempt to determine probable fundamental origins within a larger picture of greater complication, and analogies between systems in different areas, both experimental  and theoretical, in the light of  significant progress in spin glass understanding. 
}

\section{Introduction}
\label{sec:1}

The existence of macroscopic magnetism has been known since ancient times, with appreciation of its possible spontaneous microscopic origins coming from the mean-field theories of  Weiss (1907) \cite{Weiss} and Stoner (1938) \cite{Stoner}, respectively for local-moment and  itinerant ferromagnets. 
 The electrical analogue, ferroelectricity, was discovered experimentally in 1920 \cite{Valasek}. 
The subsequent recognition of antiferromagnetic and ferrimagnetic  orderings is due to N\'eel \cite{ Neel2}. In these conventional phases, as well as in many other subsequently discovered ferroic phases, the order is macroscopically periodic, as well as of lower symmetry than the corresponding higher temperature para-phases, which lack long range ferroic order.

The recognition of the existence of different dipolar-glassy behaviour in certain alloys, quasi-frozen locally but without periodic ferroic order, dates back some half a century in both magnetic and electrical scenarios. Initially it was thought `just' to represent slowing down of dynamics with reduced temperature  as experienced in conventional glasses, but interest in the magnetic  `glasses' became
more focussed 
with the observation, at the beginning of the 70s, of sharp but non-divergent  low-field magnetic susceptibility peaks as a function of temperature in {\bf{Au}}Fe alloys  \cite {CannellaMydosh}, suggesting a conceptually new type of phase transition. In combination with evidence of local spin freezing through M{\"o}ssbauer experiments and of lack of periodicity through neutron diffraction experiments, the new state became known as `spin glass'. 
Attempting to understand these observations led to theoretical modelling and novel theoretical, experimental and computational methodologies \cite{EA,SK,Parisi, Mezard} that exposed subtle new  concepts and useful applications, not only in many material systems but also in many  physically very different  complex systems/problems, such as neural networks, hard optimization, protein-folding and also  probability theory. At a model level the underlying physical origins of the behaviour are reasonably understood, although some controversies remain, and many material examples are now known; see{ \it{e.g.}} \cite{Binder-Young, Fischer-Hertz,   Mydosh1, Nishimori, Mydosh2, Kawamura, Panchenko}.

Independently, a potentially related observation  was made already in the 50s and 60s in ferroelectric alloys \cite{Smolenskii, Smolenskii2}, in the form of  peaks  in the a.c.  electrical susceptibility of the perovskite alloy $\mathrm {Pb(Mg}_{1/3}\mathrm{Nb}_{2/3}\mathrm{)O}_3$ (PMN), with significant frequency-dependence, no ferroelectricity and no change of global symmetry, at temperatures much below those of the relatively frequency-independent ferroelectric transition in the related  non-disordered compound $\mathrm {PbTi}\mathrm {O}_3$ (PT). This new behaviour was named `relaxor'.
The discovery of the relaxor behaviour in ferroelectric alloys
\footnote{We use the expression `ferroelectric alloy' to refer to alloys which exhibit ferroelectricity (or antiferroelectricity) at appropriate concentrations and low enough temperatures.}
also sparked much interest and practical application, but its fundamental origin has remained uncertain and contested. 

A third type of ferroic glass can be found in martensitic alloys, given the name `strain glass' \cite{Ren}, but this was a more recent discovery, despite the fact that practical interest in martensites  goes back to the nineteenth century.

In this chapter I shall try to relate these different types of ferroic glasses under a common conceptual umbrella, including both well-defined local moments and induced moments, within minimal modelling.
 
 \section{Experimental indications}
 \label{sec:2}
 
Before giving a theoretical discussion, it is suggestive to note some further similarities in experimental observations of different systems.

 \begin{figure*}
\includegraphics*[width=.3450\textwidth,height=.30\textwidth]{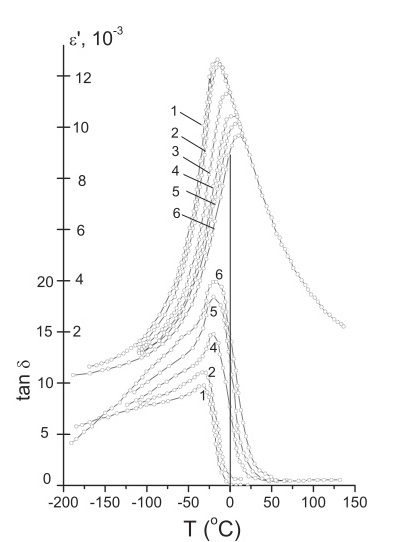}\hfill
\includegraphics*[width=.3050\textwidth,height=.30\textwidth]{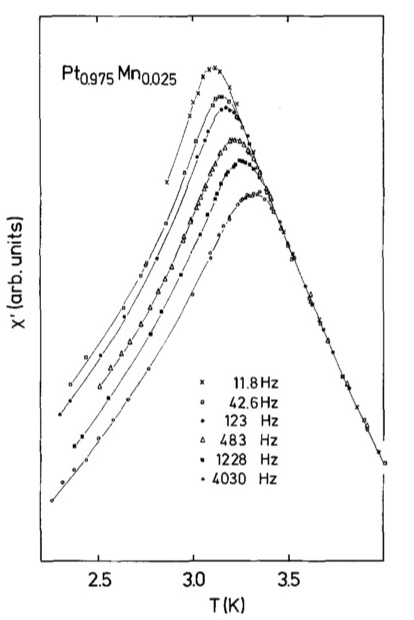}\hfill
\includegraphics*[width=.3450\textwidth,height=.30\textwidth]{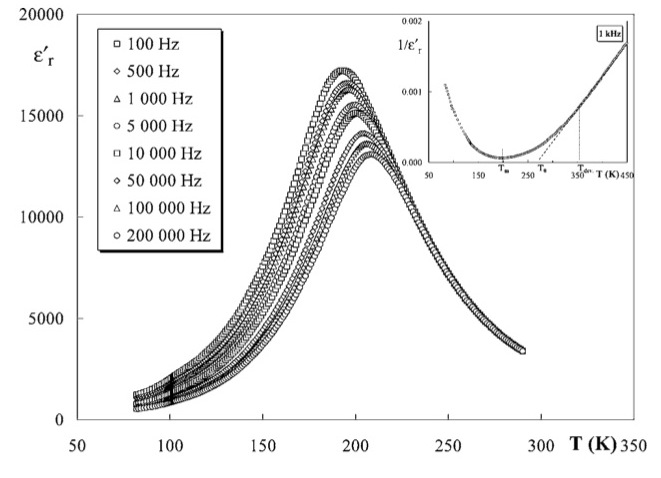}\hfill
\caption{%
AC susceptibilities; heterovalent relaxor  $\mathrm {Pb(Mg}_{1/3}\mathrm{Nb}_{2/3}\mathrm{)O}_3$ (PMN) \cite{Smolenskii}, spin glass {\bf{Pt}}Mn \cite{Wassermann}  \copyright {Springer 1983},
homovalent relaxor  $\mathrm{BaZr}_{0.35}\mathrm{Ti}_{0.65}\mathrm{O}_3$ (BZT) \cite{Maglione} \copyright{IOPP (2004) }
.}
\label{fig:ac_susceptibilities}
\end{figure*}
\begin{figure}
\hspace{10 pt.}
\includegraphics*[width=.3650\textwidth,height=.350\textwidth]{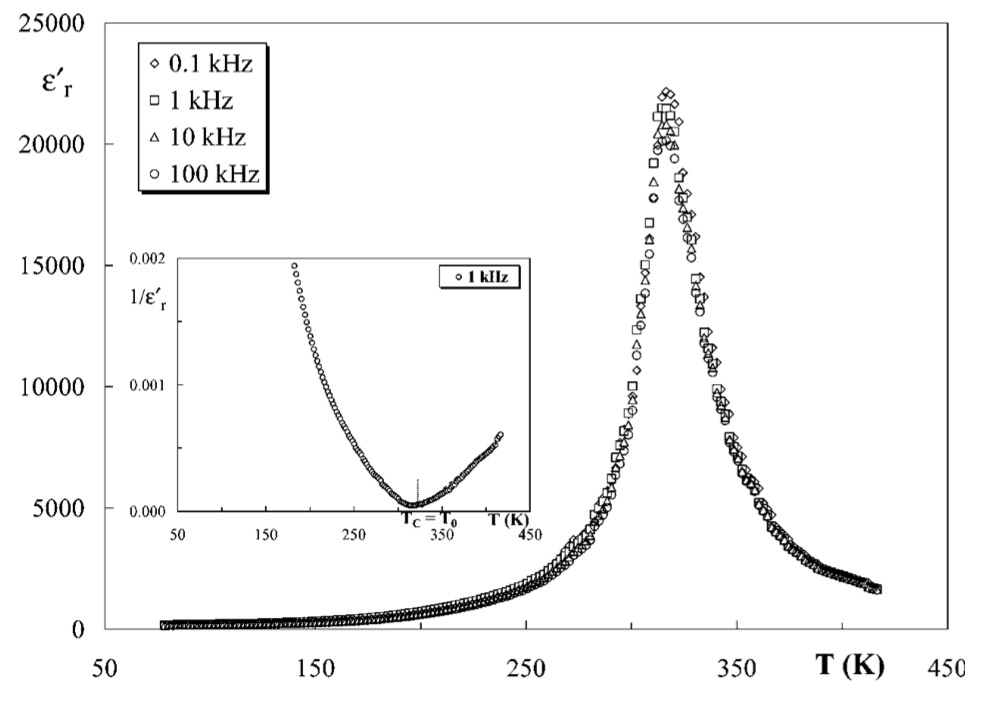}
\hfill
\includegraphics*[width=.360\textwidth,height=.35\textwidth]{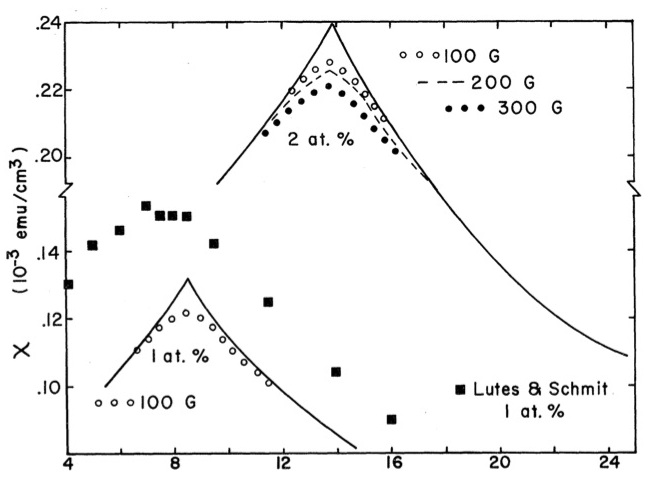}
\hspace{10 pt.}
\caption{(a) a.c.electrical susceptibility of ferroelectric $\mathrm{BaZr}_{0,2}\mathrm{Ti}_{0.8}\mathrm{O}_3$ (BZT) at several frequencies \cite{Maglione}   \copyright{IOPP (2004) }; (b) low-field magnetic susceptibility of two ${\mathrm{{\mathbf{Au}}Fe}}$ alloys under different applied fields \cite{CannellaMydosh} \copyright{APS (1972)}}
\label{Susc_ferro_BZT_AuFe}
\end{figure}

In Fig \ref{fig:ac_susceptibilities}, are shown a.c. susceptibilities (electrical or magnetic, as appropriate), of the original (heterovalent) relaxor PMN, the spin glass $\mathrm{Pt}_{1-x}\mathrm{Mn}_{x}$ at $x=0.025$, and the more recently discovered homovalent relaxor 
$\mathrm{BaZr}_{1-x}\mathrm{Ti}_{x}\mathrm{O}_3$ (BZT) at $x=0.65$. 
They are clearly very similar, with peaks indicative of transitions or strong crossovers, with strong frequency dependence, slow to respond and glassy, suggesting that similar physics is at play in these experiments. Yet they are rather different in several other aspects of their physical make-ups; both PMN and BZT are ceramic (insulating) substitutional alloys with the basic average perovskite structure $\mathrm{ABO}_3$, where A is an ion of charge 2+, B is an ion of charge 4+ and O has charge 2-, but with random substitution on the B sites; however,  in BZT the replacement B ions also have charge 4+, hence the labelling as `homovalent', while in PMN the replacement B ions have charges 2+ for Mg and 5+ for Nb, in ratio 1:2 to maintain the average charge, hence the description as `heterovalent'; { \bf{Pt}}Mn is a face centred cubic metallic alloy with magnetic moments only  on the Mn. It is thus natural to look for conceptual common links beyond normal material appearances.

For comparison/contrast, Fig \ref{Susc_ferro_BZT_AuFe}(a) shows the corresponding susceptibilities of BZT at a concentration at which the alloy is ferroelectric, demonstrating no  significant frequency dependence and hence no glassy slow response. Fig  \ref{Susc_ferro_BZT_AuFe}(b) shows the effects of even small applied fields in ${\textrm{\textbf{Au}}}{\textrm{Fe}}$, rounding the transition  but also   suggesting that it is sharp in the limit of zero applied field. One can also note that although the (normal) susceptibility diverges at a second-order ferromagnetic or ferroelectric transition, it does not diverge at spin glass or relaxor transitions, indicating that the global moment is not a primary order parameter for a spin glass or relaxor.

\begin{figure*}
\includegraphics*[width=.340\textwidth,height=.25\textwidth]{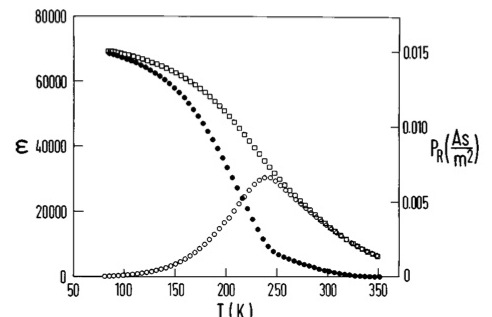}\hfill
\includegraphics*[width=.310\textwidth,height=.235\textwidth]{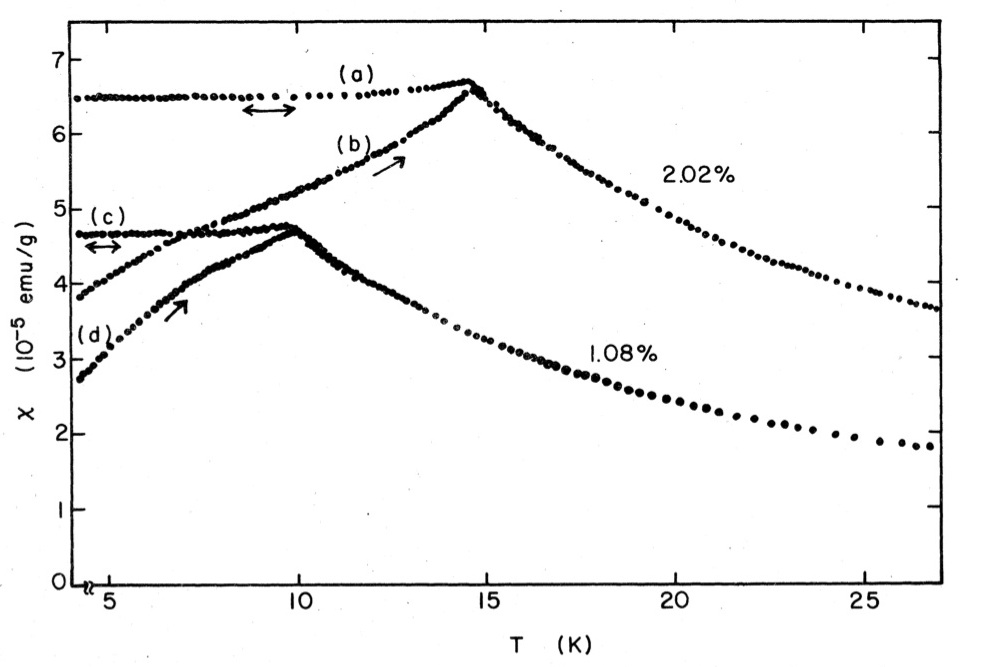}\hfill
\includegraphics*[width=.340\textwidth,height=.245\textwidth]{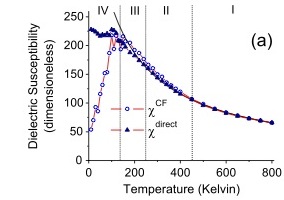}%
\caption{
Field-cooled (FC) and zero-field-cooled (ZFC) static susceptibility measurements; PMN \cite{Levstik} \copyright{APS (1998)}, spin glass {\bf{Cu}}Mn \cite{Nagata} \copyright{APS (1979)}, BZT(50:50) simulation \cite{Akbarzadeh2012} \copyright{APS (2012)}
}
\label{fig:FCZFC}
\end{figure*}

In 
Fig \ref{fig:FCZFC}
are shown for comparison examples of the of field-cooled (FC) and zero-field-cooled (ZFC) susceptibilities for the heterovalent relaxor PMN \cite{Levstik} and the spin glass {\bf{Cu}}Mn \cite{Nagata}, along with results of computer simulation of analogous measures for a model of the homovalent relaxor BZT \cite{Akbarzadeh2012}. Again there are clear similarities as the temperature is reduced through that associated with the low-frequency a.c. susceptibility peak, of the continuous separation of the two kinds of susceptibility measure, cooling in the probe field (FC) and cooling without the field and then applying the field to measure (ZFC), respectively understood as probing all thermodynamic states (FC)  and probing only accessible states (ZFC), the separation indicating the onset of a hierarchy of barriers.

\section{Spin Glasses}
\label{sec:3}

\begin{figure}
  \includegraphics[width=0.31\linewidth,height=0.35\linewidth]{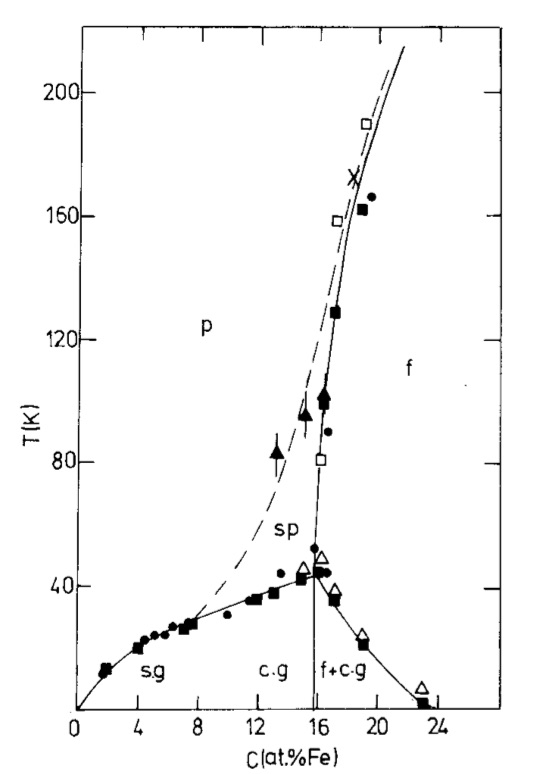}\hfill
  \includegraphics[width=0.31\linewidth,height=0.35\linewidth]{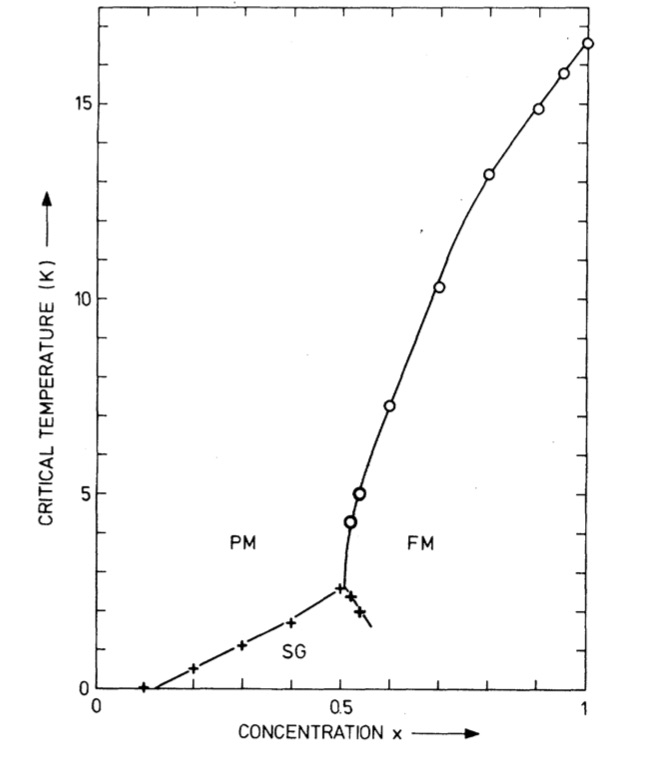}\hfill
   \includegraphics[width=0.31\linewidth,height=0.35\linewidth]{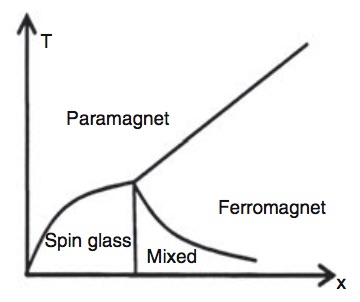}
   \caption{%
Spin glass phase diagrams; (a) metal: AuFe
 \cite{Coles_AuFe} \copyright{Taylor and Francis (1978)}, 
(b) semi-conductor:$\mathrm{Eu_{x}Sr_{1-x}S}$,
\cite{Maletta}\copyright{APS (1979)},
(c)
    SK model with mean and variance of exchange distribution both scaling with concentration $x$.
 \cite{Sherrington_Glassy} \copyright{Springer (2012)}
}%
\label{fig:phase_diagrams}
\end{figure}

The canonical spin glasses, such as {\bf{Au}}Fe and {\bf{Cu}}Mn, involve non-magnetic hosts, Au and Cu, and a finite concentration of local-moment-bearing substitutions, Fe and Mn. Paramagnetic at high temperatures, they exhibit spin glass behaviour beneath critical temperatures at lower (but finite) concentrations of magnetic ions. A similar behaviour is also found in many other systems, both metals and insulators; see 
{\it{e.g.}} Fig  {\ref{fig:phase_diagrams}}.

      To model the cooperative magnetic behaviour one typically expresses the Hamiltonian as 
\begin{equation}
H_{CSG}=-\sum_{(ij)(Mag)}  {J}({\textit{\textbf{R}}}_{ij}){\textit{\textbf{S}}}_{i}.{\textit{\textbf{S}}}_{j}
\label{eq:Hcsg}
\end{equation}
where the ${\textbf{S}}_{i}$ are localized spins, of fixed length but variable direction, located on the magnetic ions, 
${J}({\textit{\textbf{R}}})$ is a translationally-invariant but
spatially-frustrated `exchange interaction' and the sum is over pairs of sites occupied by magnetic atoms. 

For the canonical metallic systems such as {\bf{Au}}Fe and {\bf{Cu}}Mn, the effective interaction between the magnetic ions is carried by the conduction electrons via the s-d coupling, resulting in the RKKY form 
\begin{equation}
{J}({\textit{\textbf{R}}}_{ij}) ={\mathcal J}^{2}{\chi_{ij}}
\label{eq:RKKY}
\end{equation}
where $\mathcal{J}$ is the coupling strength between the conduction electron spin (${\textit{\textbf{s}}}_{i}$) and the local moment spin (${\textit{\textbf{S}}}_{i}$) 
 and $\chi_{ij}$ is the conduction band susceptibility 
  between sites $i$ and $j$. $\chi_{ij}$ 
  oscillates in sign with separation $R_{ij}$, with wavevector $2k_F$ where $k_F$ is the Fermi wavevector, and (in 3 dimensions) also decays in magnitude as $R^{-3}$. The oscillation in sign results in a competition in ordering tendencies of the spins, now known as `frustration' \cite{Toulouse}, while the randomness of occupation of lattice sites by magnetic ions provides quenched disorder and inhomogeneity of local environments. 
 
The RKKY interacting metal systems are, however, just one experimental example of the combination of frustration and disorder leading to spin glass behaviour. In the second example of Fig {\ref{fig:phase_diagrams} the material is semiconducting, the spins are on the Eu and their interaction arises from shorter-range superexchange, with frustration due to competition between nearest-neighbour and antiferromagnetic next-neighbour interactions. 
 
It is now well-established that the combination of frustration and quenched disorder are the key ingredients for spin glass behaviour. This realisation by Edwards and Anderson (EA)  \cite{EA} led them to to suggest in 1975 an alternative model for potentially easier but conceptually equivalent theoretical study, along with further new conceptualization and methods of analysis that ignited theoretical excitement.  In their model every site is occupied by a magnetic spin but their interactions are chosen randomly and quenched:
\begin{equation}
H_{EA} = -\sum_{(ij)} J_{ij}{\textit{\textbf{S}}}_{i}.{\textit{\textbf{S}}}_{j} 
\label{H_EA},
\end{equation}
where the  $J_{ij}$ are chosen randomly from a symmetric distribution of mean zero,
ensuring that no conventional periodic order is possible. 

Through novel and innovative analysis EA demonstrated the existence  of a new phase with random spin-freezing. 
They noted that a relevant order parameter to test for spin freezing, independent of overall periodic order, is 
\begin{equation}
q_{EA}=\lim_{\tau  \rightarrow \infty}\overline{S_{i}(t)S_{i}(t+\tau)}
\label{q_time},
\end{equation}
where the overbar refers to an average over sites $i$ and times $t$, or, equivalently  
\begin{equation}
q_{EA}= \overline{  {\langle S_{i}  \rangle}^2   }
\label{q_thermo},
\end{equation}
where the $\langle . \rangle$ brackets refer to a thermodynamic average and the overbar to a site/disorder average. Thus, `amorphous' spin freezing without ferromagnetism is signalled by non-zero $q_{EA}$ but zero overall magnetization $m$, as given by 
\begin{equation}
m= \overline{  {\langle S_{i}  \rangle_{i}}   }
\label{m_thermo}.
\end{equation}

The EA model has become an important paradigm in further theoretical  study. It is normally considered as having only nearest neighbour interactions on a simple cubic (or hypercubic) lattice.  Computer simulations have demonstrated that it captures key features of real systems. An extension  to allow for competition of the spin glass phase with ferromagnetism by allowing a finite mean $J_{0}$ to the interaction distribution, of standard deviation $J$, by Sherrington and Southern (SS)  \cite{SS}, 
showed that when  $J_{0}/J$ is large enough the low temperature state is a ferromagnet, while for smaller $J_{0}/J$, beneath a critical value, the low temperature state is spin glass. 

The EA model with finite interaction range is not exactly soluble. However, an extension in which the distribution from which the interactions are drawn  is the same for all pairs of sites, independently of their separation, the Sherrington-Kirkpatrick (SK) model  \cite{SK}, is soluble, althought its solution is very subtle, requiring a description beyond that of a single simple order parameter \cite{Parisi},  and has exposed several unexpected but interesting  features and concepts \cite{Mezard}. Its solution clearly demonstrates the existence of phase transition to a glassy phase,  even in an applied field, and also that its spin glass phase has a complex structure with a hierarchy of metastable states and chaotic evolution under change of global controls (such as temperature).  It has stimulated much further study in many other range-free random problem scenarios. However, there remains  controversy about whether all the conceptual results of the SK model studies apply to finite-ranged systems, especially those related to so-called replica-symmetry-breaking \cite{Parisi} and to whether a phase transition still persists in an applied field.

\subsection{Simulations}
Computer simulations of model systems 
have played an important role in determining whether true phase transitions exist also in systems with range-dependent interactions, using their ability to measure directly 
observables which are not readily accessible to conventional experimentation, such as 
the spin glass order parameter $q_{EA}$ and a related spin glass susceptibility,  as well as the more conventional measures such as the ferromagnetic order parameter $m$.

 The existence of true phase transitions can be tested through sophisticated simulation studies,
especially through the use of finite-size scaling and Binder plots\cite{Binder}.
These studies have provided  clear demonstrations of spin glass phase transitions in several interesting situations, {\it{e.g.}} as illustrated in  Fig {\ref{fig:crossover plots}} for three examples; (i)  spin-glass correlations in the SK model with zero mean exchange \cite{Bhatt-Young}, (ii) a nearest neighbour Ising EA model in dimensions 3 (again with zero mean exchage) \cite{Katzgraber-Young} and (iii) a longer-range dipolar model emulating 
 ${\mathrm{LiHo}_{x}}      \mathrm{Y_{(1-x)}F_{4}}$ at $x=0.001$, a concentration at which the system is a spin glass
 \cite{Katzgraber_PRX} \footnote{The phase transitions are demonstrated by the crossing of appropriate correlation measures for systems of different sizes, with scaling plots providing further confirmations and exponents.}. Corresponding plots for ordinary magnetic correlations in these systems do not show crossings, indicating the absence of a  ferromagnetic transition. The combination of these two results, crossing of the size-normalized spin glass correlation lengths together with the lack of crossing of the normal magnetic correlation lengths, lead to the deduction of a true spin glass phase transition at the crossing temperature. 
 \begin{figure*}
\includegraphics*[width=.320\textwidth, height=.30\textwidth]{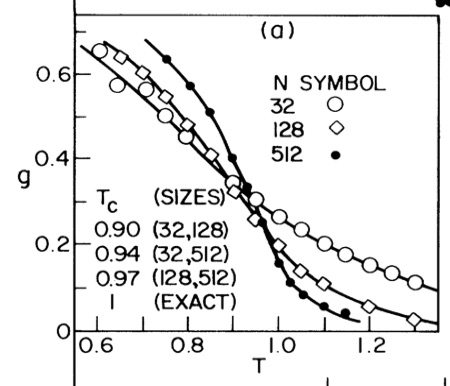}\hfill
\includegraphics*[width=.320\textwidth,height=.30\textwidth]{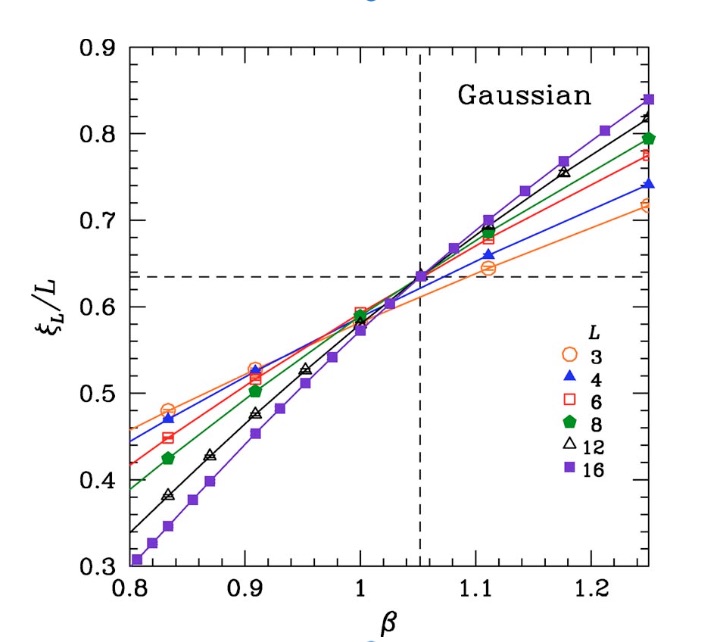}\hfill
\includegraphics*[width=.320\textwidth,height=.30\textwidth]{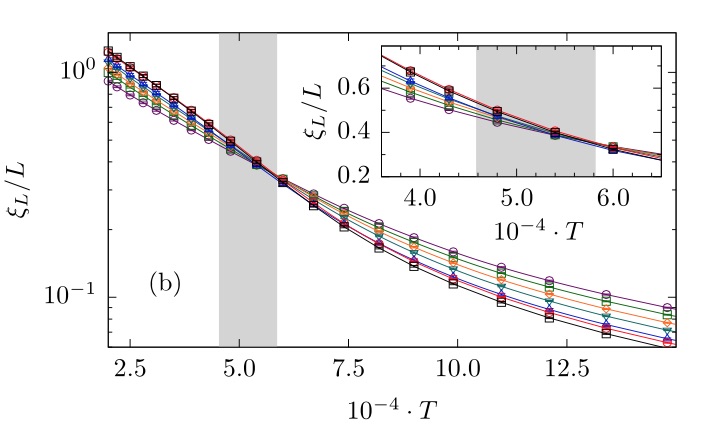}
\caption{Spin glass correlation plots
for different sample sizes, with  crossovers at phase transition temperatures, for (a) SK model \cite{Bhatt-Young} \copyright{APS (1984)}, (b) three-dimensional Ising EA  \cite{Katzgraber-Young} \copyright{APS 2006} and (c) ${\textrm{{LiHo}}_{x}\mathrm{Y}_{(1-x)}\mathrm{F_{4}}}$;$x=0.001$  \cite{Katzgraber_PRX}
}
\label{fig:crossover plots}
\end{figure*}

\subsection{Soft spins}

In analytic studies of  spin glasses the Hamiltonian is often re-expressed using continuously-valued `spin fields' $\{\bf{\phi\}}$ in place of the fixed length spins $\{{\textit{{\textbf{S}\}}}}$. Within a 
simplification  to one-dimensional (Ising) spins
$H_{EA}$ becomes

\begin{equation}
H_{EA{\mathrm{cont}}} = 
\sum_{i}( r{{\phi}_{i}}^2 + u {{\phi}_{i}}^4) - \sum_{(ij)}
 J_{ij}
 \phi_{i}
\phi_{j}
\label{H_EA_cont},
\end{equation}
while the analogue for site-disorder is

\begin{equation}
H_{CSG{\mathrm{cont}}}=
 \sum_{i}
  ( r_{i} \phi_{i}^2 
+ u_{i} \phi_{i}^4)  
- \sum_{(ij}  {J}
({\textit{\textbf{R}}}_{ij})
\phi_{i}\phi_{j}
.
\label{eq:Hcsg_cont}
\end{equation}
The full hard-spin Ising case ($S=\pm {1}$) results from taking the limits 
\begin{equation}
r \rightarrow- \infty, u \rightarrow \infty, r/2u \rightarrow -1
\label{Ising_limit}.
\end{equation}
The sums are taken over only magnetic sites or, equivalently, the non-magnetic sites can be emulated by taking $r_{i} \rightarrow \infty$ on those sites .
Note, however, that if some $r_{i}$ are negative but finite then for those sites to displace there needs to be a sufficient binding energy from the interaction term to overcome the local quadratic  penalty for displacements, otherwise the ground state would have $\phi =0$. Such bootstrapping is referred to as `induced moment'. Note that the resultant order will depend on the character of the interactions and can be either  globally periodic, including ferromagnetic, or spin glass in a system with sufficient disorder and frustration.
 
Early experimental indications of induced moment  spin glass behaviour were found in  the alloys {\bf{Y}}Tb and {\bf{Sc}}Tb \cite{Sarkissian-Coles}, in which crystal field effects lead to a singlet ground state for isolated Tb ions. For Tb concentrations less than a small but finite percentage the ground state is non-magnetic, in contrast to the corresponding alloys with non-singlet ground state Gd in place of Tb, in which the spin glass state continues to the lowest finite conentrations.

  A simple extension of the EA model  exhibiting induced moment spin glass behaviour was introduced in 1977 by Ghatak and Sherrington (GS) \cite{Ghatak};
\begin{equation}
H_{GS} =- \sum_{i}DS_{i}^2 -\sum_{(ij)} J_{ij}{\textit{\textbf{S}}}_{i}.{\textit{\textbf{S}}}_{j}
\label{H_GS}
\end{equation}
with the S taking values  $S=0,\pm1$ and the $\{J\}$ again drawn randomly from a distribution of mean zero. For D less than a critical (negative) value $D_c$  there is only a paramagnetic phase, while above there is an induced-moment spin glass phase.

\section{Polar glasses and relaxors}
\label{bsec:3}

Ferroelectric systems are often categorized as being of polar/`order-disorder' type or `displacive' type. 

In the former one envisages local electric dipolar moments well formed (but not cooperatively ordered)  already in the paraelectric phase above the macroscopic ordering transition to ferroelectricity (or, if energetically preferable, to another periodic phase), in close analogy with local moment magnetism. 
Correspondingly,  alloys with  sufficient dilution of local moment units by neutral ones, together with frustrated interactions, can lead to close analogies of conventional local moment spin glasses 
\cite{Hochli}
\cite{Vugmeister}. Extensions of spin glass modelling and analysis  have also been developed for systems characterised by the interaction of higher-order local moments   \cite{Goldbart} \cite{Sherrington-Japan} \cite{Binder-Reger}.

By contrast, in displacive ferroelectrics there are no long-lived electric moments above the transition to ferroelectricity and the charged ions fluctuate around a mean lattice structure with no overall electric moment.  Rather, in such ferroelectrics, as the temperature is lowered beneath the transition temperature
the time-averaged positions of  charged ions 
displace collectively   in such a manner as to yield overall ferroelectricity. The transition to ferroelectric is  typically accompanied by a change in global symmetry but the ferroelectricity itself  is caused by a relative distorsion of positively and negatively charged ions within the unit cells, yielding electric moments.
Unless pre-empted by a first order transition, the susceptibility diverges at the transition.
However, not all candidate systems with the same ionic charges and higher temperature structures do exhibit cooperative ordering; for example $\mathrm {BaTiO}_3$ (BT) is a displacive ferroelectric while $\mathrm {BaZrO}_3$ (BZ) is not. An energetic advantage of distorsion is needed. 

Displacive ferroelectrics can be modelled by considering the displacements of the ions as variables governed by Hamiltonians  including local costs, the (non-local) effects of interactions between displacements at different sites and  the effects of charges on different sites, with coefficients calculable by first-principles methods, followed by computer simulations at finite temperatures. 

A detailed 
first-principles theoretical/computational 
study of $\mathrm {BaTiO}_3$ was given in \cite{Zhong} and demonstrated the ferroelectric transition; see also \cite{King-Smith}. However, the conceptual principles can be seen already from a simplified model allowing only for  one-dimensional displacements of the most polarizable ions:
\begin{equation}
H_{R}=\sum_{i} 
\{
\kappa u_{i}^2 +\lambda u_{i}^4
\}
+ {\sum_{ij} J_{ij}  u_{i}u_{j}}
\label{siple ferroelectric}
\end{equation}
where the $u_{i}$  are the displacements of the ions
 at sites $\{{i}\}$, the first (single-site) term describes the local energy costs of displacements and the last term represents the interaction energy. Clearly, this has a similar form to Eqn. {(\ref{H_EA_cont})} and can yield an induced moment (displaced $u \neq 0$) ground state if the energy minimising  gain from the interaction term can overcome the local cost from the $\kappa$ term, with a corresponding transition at a higher temperature. 
For $\kappa$ close to zero one expects features of  both displacive and order-disorder behaviour, reducing $\kappa$  making it more order-disorder-like.

 \begin{figure}[h]
\sidecaption[h]
\includegraphics[width=.60\textwidth,height=.320\textwidth]{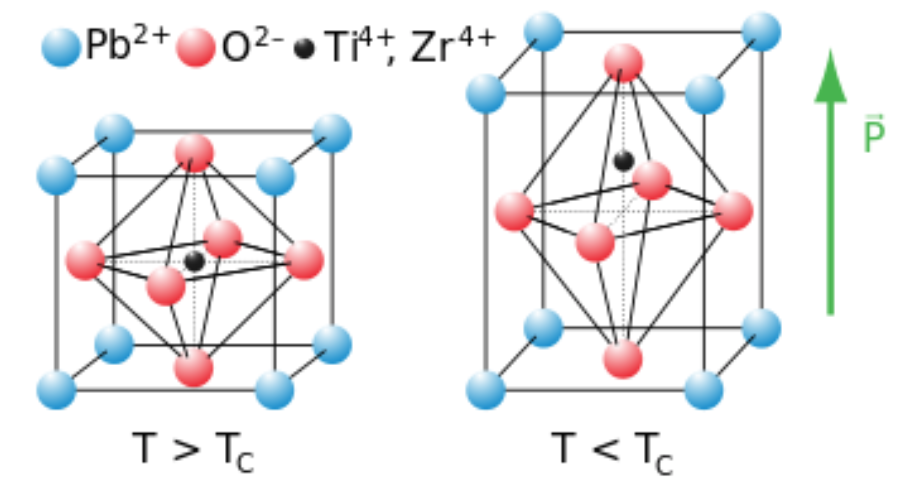}
\caption{Unit cell structure of $\mathrm {PbTiO}_3$ above and below the ferroelectric transition temperature.  $\mathrm {BaTiO}_3$ is similar, but with smaller tetrahedral stretching. \vspace{1 cm}
}%
\label{fig:perovskite}
\end{figure}

Here, however, the main interest is in alloys. In particular, we shall concentrate on alloys of underlying perovskite structure $\mathrm{ABO}_3$ with substiutional disorder on the B sites. This disorder can be either homovalent, for which the ions on the B-sites all have the same 4+ charge as the template, or heterovalent, for which the B-ions have  different charges but with the average charge of 4+.

\subsection{Homovalent relaxors}
The homovalent alloy  $\mathrm {Ba}(\mathrm{Zr}_{1-x}\mathrm{Ti}_{x}\mathrm{)O}_3$ (BZT) exhibits ferroelectricity at higher $x>x_{c1}$, only paraelectricity for $x<x_{c2}$, with relaxor behaviour in between \cite{Maiti}. The present author has argued that the relaxor state of BZT is essentially an induced moment spin glass \cite{DS_BZT}. The susceptibility measured in BZT in a relaxor region of the concentration $x$ is shown in Fig \ref{fig:ac_susceptibilities}.

  Here we shall use only a simplified model to illustrate the probable origin of the relaxor behaviour observed in BZT at intermediate concentrations \cite{Kleemann-Miga} .
We 
note that at the para- to ferro-electric transition, while the overall lattice structure stretches from cubic to tetrahedral, the B-site ions displace from the symmetric lattice positions; see Fig {\ref{fig:perovskite}}, yielding ferroelectricity.. Also, it is observed that in the relaxor state the overall average lattice structure remains cubic.  Hence,
while all the ion locations are, in principle, variable, we shall initially ignore any A and O site displacements, coupling to global strain and change in global lattice structure  and
concentrate on the deviations of the B-site ions  from their locations on the pure perovskite $\mathrm{ABO_3}$ lattice, 
using 
\begin{equation}
H_{R}=\sum_{i} 
\{\kappa_{i}| \textit{\textbf{u}}_{i} |^2
+\lambda_{i}|\textit{\textbf{u}}|^4    
+ \gamma_{i}  
{{(u_{ix}^{2}u_{iy}^{2}
 + u_{iy}^{2}u_{iz}^{2} +u_{iz}^{2}u_{ix}^{2})  }}
\}
+ {\sum_{(ij)}\sum_{\alpha \beta}J_{ij}^{\alpha  \beta}      u_{i\alpha}u_{j\beta}}
\label{eq:Displacive}
\end{equation}
where the $\{ \textit{\textbf{u}}_{i}\}$  are the displacements of the ions at B-sites $\{{i}\}$, the first (single-site) term describes the local energy costs of displacements, with the $\kappa,   \lambda$ and $\gamma$ coefficients depending upon the types of atoms at those sites, and the last term represents the interaction energy,  involving a short-range contribution due to (quantum mechanical) electronic interference of neighbouring pairs of B ions, long-range Coulomb interactions and effective interactions via the ions on A and O sites. 
This is immediately recognisable as a vector analogue of Eqn.  (\ref{H_EA_cont}), also allowing for anisotropy. 

For $\kappa$ positive the ground state will have $u=0$ if the interaction strength is insufficient to overcome it. This appears to be  the case for Zr in $\mathrm {BaZrO}_3$, which is everywhere paraelectric, while for Ti $\kappa$ is smaller  and $\mathrm {BaTiO}_3$ is ferroelectric. Empirically, for a pure system at finite temperature, one can consider  Eqn ({\ref{eq:Displacive}}) to represent instead an effective Landau free energy with temperature-dependent  coefficients, 
\begin{equation}
\kappa=\kappa_{c} + a (T-T_c) + {\mathcal{O}}( T-T_c)^2
\label{eq:Landau}
\end{equation}
where $\kappa_{c}>0$ is the critical value at which the energy cost from the local harmonic term equals the maximum energy gain from the interaction and $T_c$ is the transition temperature.

For alloys such as BZT one can model as in Eqn. ({\ref{eq:Displacive}}) but with different $\kappa, \lambda$ and $\gamma$ on Zr and Ti sites. Given that $\kappa^{Zr}$ is too great to allow order in BZ,  the situation is  analogous to that in Eqn. ({\ref{eq:Hcsg_cont}}), albeit without the extremes of coefficients and with $r$  rather positive on Zr sites, reminiscent of the non-magnetic sites in conventional spin glasses but allowing for some paraelectric induction.

Dipolar interactions are frustrated, as well as long-ranged, well-known to lead to several different magnetic phases in different structures and in combination with different extra shorter-range interactions \cite{Luttinger_Tisza}; {\it{e.g.}} a simple cubic Ising dipolar system has an antiferromagnetic ground state, while the
tetragonal $\textrm{LiHoF}_4$ is ferromagnetic at low temperature. It is also known from experiment and from computational studies of hard-spin dipolar models that site-dilution of dipolar sites can lead to spin glass phases  in such systems; see the first two sub-figures of Fig \ref{fig:dipolar_glasses}  \cite{Fernandez} \cite{Andresen_PRX}. 

Hence it seems reasonable to anticipate a corresponding soft pseudo-spin glass phase in homovalently-diluted frustrated ferroelectics in appropriate parameter regions and for the observed relaxor state in BZT to be a manifestation of such a phase,  the pseudo-spins being the local dipoles induced by displacement of the charged B-ions. Computer simulations of   $\mathrm {Ba}(\mathrm{Zr}_{0.5}\mathrm{Ti}_{0.5}\mathrm{)O}_3$ (50:50 BZT) \cite{Akbarzadeh2012} demonstrate such behaviour; see Fig \ref{fig:FCZFC}.  

Note, however, that for any ordered phase the binding energy from the interaction term must be sufficient to overcome the cost of any positive $\kappa$. Hence the  phase diagram for soft-spin versions of the models of \cite{Fernandez} \cite{Andresen_PRX} would be expected to correspond to lowering the phase transition lines shown in the first two sub-figures of Fig \ref{fig:dipolar_glasses} by an amount of order $\kappa^{Ti}$, thereby yielding a phase diagram as indicated schematically in the third sub-figure\footnote{ Conceptually, at the simplest level, the Zr ions are analogues of the non-magnetic atoms in conventional local moment spin glasses ({\it{e.g.}} Cu or Au in {\bf{Cu}}Mn and {\bf{Au}}Fe), although in fact they should be paraelectrically displaced a small amount by interaction with the displaced Ti ions.}, in qualitative accord with observations \cite{Maglione}\cite{Maiti}\cite{Kleemann-Miga}.

\begin{figure*}
\includegraphics*[width=.345\textwidth,height=.25\textwidth]{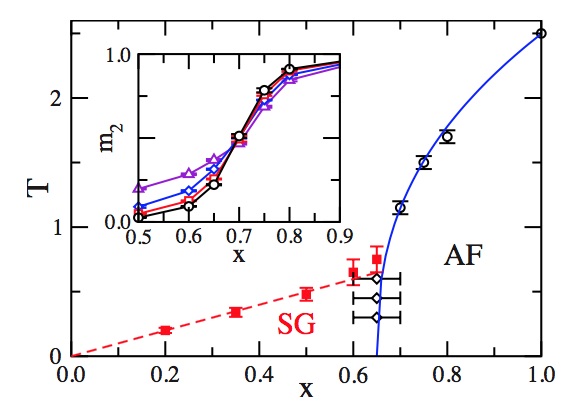}\hfill
\includegraphics*[width=.315\textwidth,height=.235\textwidth]{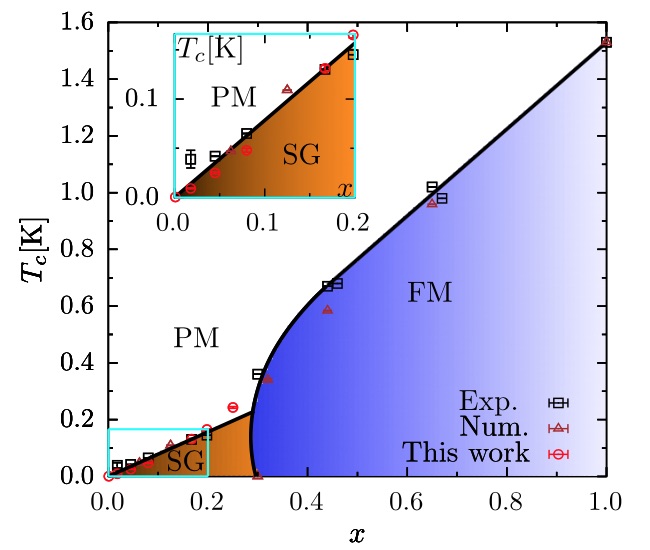}\hfill
\includegraphics*[width=.315\textwidth,height=.255\textwidth]{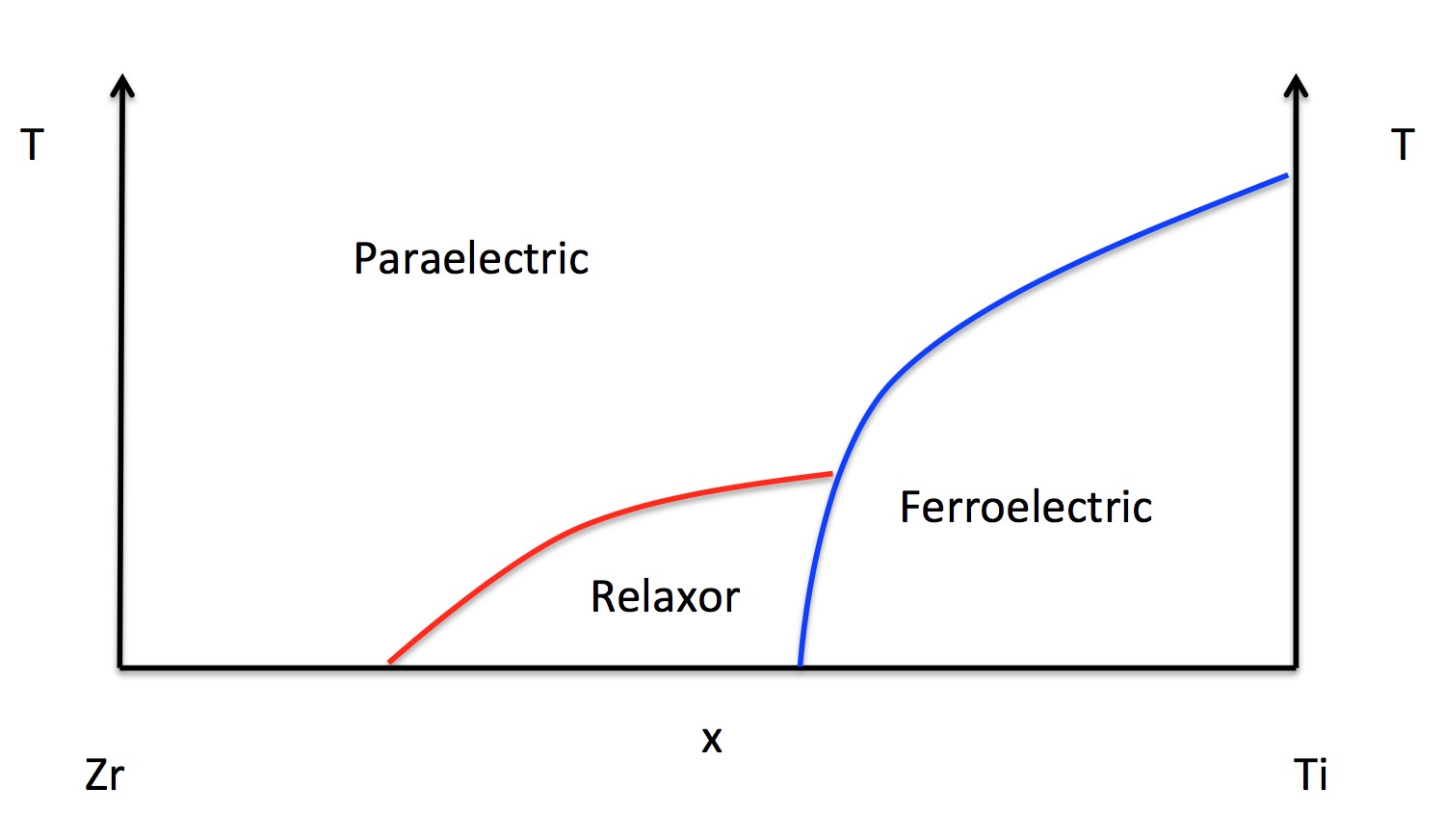}%
\caption{
Phase diagrams:  Computer simulations of site-diluted dipolar Ising models (a) on a simple cubic lattice
 \cite{Fernandez} \copyright{APS (2010)} , 
(b) on a tetragonal lattice with also short-range antiferromagnetic interaction, with parameters based on on ${\mathrm{LiHo}_{x}\mathrm{Y}_{(1-x)}F_{4}}$
\cite{Andresen_PRX}, and
(c) schematic speculation for BZT}
\label{fig:dipolar_glasses}
\end{figure*}

In the model considerations above we have ignored any possible change in the basic cubic lattice structure. This is in accord with observations for relaxors. However, in para- to ferro-electric transitions there are normally observed changes in the global average lattice structure, for example in  $\mathrm {BaTiO}_3$ and  $\mathrm {PbTiO}_3$ to a tetragonal structure. This is a consequence of inclusion of global strain coupling which we have not included explicitly; see \cite{Zhong}. It will affect the relative energetic preferences for the ferroelectricity and relaxor and hence transition compositions at phase boundaries separating them, but the present author believes it does not affect the conceptual principles given above for the existence of pseudo spin glasses and experiment shows no change in global symmetry at the relaxor transition\footnote{The absence of a global strain in the relaxor state can be attributed to the lack of an overall global moment.}.
 We have also not considered the other constituent elements explicitly, only  assumed inclusion of their contributions implicitly via the effective interactions between B-ions. Such extra effective interaction contributions will depend upon the ions on the A sites and is presumably at least part of the reason why $\mathrm{BaZr}_{1-x}\mathrm{Ti}_{x}\mathrm{O}_3$ (BZT) has a relaxor phase but $\mathrm{PbZr}_{1-x}\mathrm{Ti}_{x}\mathrm{O}_3$  (PZT) appears not to have one; while the direct B-B interactions should be similar in both alloys, the indirect interaction via the A ions will be different, with that for A=Pb  more strongly ferroelectric than that for A=Ba. In fact. both experiment \cite{Hewat}\cite{Shirane}and theoretical calculations \cite{King-Smith} show that while the Ti displacements in $\mathrm {BaTiO}_3$ are much greater than those of the Ba ions, in $\mathrm {PbTiO}_3$ the situation is almost inverted, the Pb displacements being greater than those of the Ti ions. Hence, in the Pb-based systems ideally one should  include the Pb (and O) displacements explicitly in the Hamiltonian.  However, the combination of frustration and disorder  should continue to allow for the possibility of a spin-glass-like relaxor state, albeit that it may not be a preferred one in PZT.

\subsection{Heterovalent relaxors}

The original classic relaxor PMN is heterovalent, the B-site 4+ ions of the $\mathrm{ABO}_3$ template being replaced  by Mg 2+ ions and Nb 5+ ions 
 in the ratio 1:2.  Below we attempt to move conceptually towards a possible understanding  in the light of the observations above, albeit in a discussion that  is at some variance with convention.

Let us first consider in terms of the basic Hamiltonian of Eqn (\ref{eq:Displacive}) but now with account needing to be taken of the fact that the B-ions are of different charges and hence that $J_{ij}$ depends on the particular ions at $i$ and $j$ and not simply on their separation. Allowing also for different types of A ions we shall refer to this Hamiltonian as $H^{1}_{AMN}$. Let us also introduce a corresponding  Hamiltonian $H^{1}_{AM^{*}N^{*}}$ for a fictitous material AM*N* in which the Mg++ and Nb+++++ of AMN are replaced by fictitous ions Mg*++++ and Nb*++++ which have the same properties as Mg++ and Nb+++++ except for their charges, which are ++++ as in the standard $\mathrm{ABO}_3$ template. We next note that Mg++ has an ionic radius similar to that of  Zr++++  and hence can be expected to have a similar largish  $\kappa$, while Nb+++++ and Ti++++ also have similar but smaller ionic radii, suggesting similar $\kappa$ and likelihood to displace. We shall assume that the B-ion replacement is random.  
Consequently, one might initially expect that AM*N* would have a similar phase structure to AZT at the same relative concentrations of 1:2. This would suggest that BM*N* would be a relaxor, or close to a boundary between ferroelectric and relaxor, while PM*N* would be a ferroelectric.  

Hence the observation that PMN appears to show the same sort of relaxor behaviour as BZT indicates that the difference between  $H^{1}_{PMN}$ and  $H^{1}_{PM^{*}N^{*}}$ is important in stabilising the relaxor phase in PMN. This difference is given by
\begin{equation}
H^{1}_{PMN} = H^{1}_{PM^{*}N^{*}} 
+V_{Coulomb}(Z_{i},Z_{j}, R_{ij})
 -V_{Coulomb}
 (Z^{0}_{i},Z^{0}_{j}, R_{ij})
\label{eq:PMN1}
\end{equation}
where 
$V_{Coulomb}(\tilde{Z}_{i},\tilde{Z}_{j}, \tilde{R}_{ij})$ is the Coulomb energy associated with charges $\tilde{Z}_{i}$ and $\tilde{Z}_{j}$ separated by a distance $\tilde{R}_{ij}$,
the $\{Z_{i}\}$ are the actual charges at sites $\{i\}$ while $Z^{0}_{i}$ is the charge at site $i$ accounted for in PM*N* ({
\it{i.e.}}  for B-sites, $Z^{0} =4+$ , for A-sites  $Z^{0} =2+$  and for O-sites $Z^{0} =2-$), and
\begin{equation}
R_{ij}=| {\mathbf{R}}^{0}_{i} +{\mathbf{u}}_{i} -{\mathbf{R}}^{0}_{j} - {\mathbf{u}}_{j} |.
\end{equation}
Expanding, the perturbation component  compared with PM*N* includes terms both linear  and bilinear in the displacements \cite{DS_Stirin}.
The coefficients of the linear terms can be viewed as effective fields and the bilinear terms as effective extra interactions. The effective fields at any site $i$ depend upon the types of ions on all sites $j \neq i$. Given that the B-site interactions are (quasi-)random, so are the effective fields.

Let us concentrate now on the possible effects of including the random fields, which have been considered as driving forces for relaxor behaviour in PMN, particularly since the work of \cite{Westfal}; for  more recent discussion see \cite{Kleemann2} \cite{Kleemann_this_book}.

Microscopically random magnetic fields are difficult to produce so there is little  experiment to compare directly with in magnetic systems; rather, diluted antiferromagnets have been studied in uniform fields, emulating ferromagnets in random uniaxial $\pm h$ fields; in the context of relaxor analogies see \cite{Cowley}.

The problem of the statistical physics of a system controlled by the Hamiltonian of 
Eqn.({\ref{eq:PMN1}})
is not soluble exactly 
and raises many questions. One relates to whether a system with a spin-glass transition in the absence of applied fields should continue to exhibit a sharp transition in the presence of such field(s).
It is accepted that the range-free SK model (with spins of any dimension) has an ergodic-non-ergodic spin glass transition even in the presence of uniform or randomly chosen local fields  \cite{Almeida-Thouless}
\cite{Sharma-Young}
\cite{GT}
\cite{Cragg}. 
On the other hand, there remains controversy about the effects of fields in short-range spin glasses, with many authors arguing that they destroy sharp spin glass transitions, on the basis  of both theoretical arguments and computer simulations,
but still without a clear accepted answer \cite{Katzgraber-Young2} \cite{Larson} \cite{Janus_field} \cite{Singh-Young}. 
Most computer simulations have been performed on Ising EA-like model systems with interactions drawn randomly from symmetric (zero-mean) distributions, whereas in the relaxor systems there are biases in the overall effective interactions, as demonstrated by the existence of 
ferroelectric phases in appropriate concentration regimes. 
Most of the simulated models 
have also had short-range nearest neighbour interactions or  are on one-dimensional structures employed to emulate short-range systems in different dimensions.

It is generally accepted that random fields have a detrimental effect on tendencies for ferromagnetism and that for sufficient  strength they suppress ferromagnetism. Thus, the effective random fields  in PMN can be expected to act to reduce  the ferroelectric tendency anticipated above in PM*N*. An approximate Ising analogue of  interactions in PMN has, in fact, been studied in computer simulations of Ising spins based on magnetic (Ho)  sites of the diluted alloy ${\mathrm{LiHo}_{x}\mathrm{Y}_{(1-x)}F_{4}}$ \cite{Andresen2013}
 and of an  EA model with non-zero mean exchange   \cite{Andresen2017}, each in the presence of random fields; see Fig \ref{fig:random_fields}.
 These simulations also indicate  what their authors call a `quasi-spin-glass' 
   in not-too-large random fields, including 
   the existence of parameter regions where the quasi-spin-glass is preferred to the ferromagnet in sufficient finite random fields, 
even though at lower random fields the opposite is the case and for higher fields the system is paramagnetic.
   It is tempting to wonder whether PMN might  lie in such a region, hence relaxor.  However, 
    more study is needed, particularly of the transition/crossover from paramagnet to (quasi-)spin glass; currently there is no computational study indicating a  sharp transition from paraelectric to relaxor, as sugested by extrapolation to zero frequency of the a.c. susceptibility observed experimentally in PMN. 

\begin{figure*}
\includegraphics*[width=.430\textwidth,height=.32\textwidth]{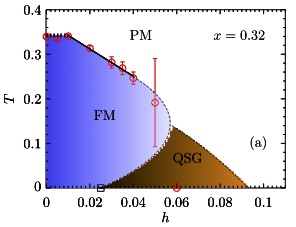}\hfill
\includegraphics*[width=.440\textwidth,height=.335\textwidth]{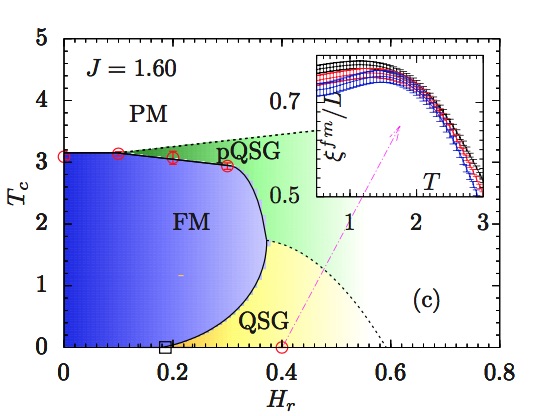}
\caption{
Phase diagrams:  Computer simulations of 
(a) diluted Ising model based on ${\mathrm{LiHo}_{x}\mathrm{Y}_{(1-x)}F_{4}}$ plus Gaussian distributed quenched random fields of standard deviation $h$ \cite{Andresen2013} \copyright{APS (2013)}
(b) n.n. EA/SS Ising model with finite mean $J_{0}=1$ and variance $J$ with Gaussian-distributed random fields of standard deviation $H_{r}$.
\cite{Andresen2017}.}
\label{fig:random_fields}
\end{figure*}

 We also note that
 computer simulational study of an Ising dipolar system on a simple cubic lattice has indicated \footnote{via a study of the size dependence of the spin glass correlation length, showing no Binder crossover.} that the sharp spin glass transition seen in zero applied field is removed in a uniform field \cite{Fernandez_H}.
Note, however, that random $\pm {h}$ fields cannot be simply gauged away into a uniform field $h$ in systems with non-zero mean exchange, as they can in the usually-studied models with zero mean. Furthermore, the magnitudes of effective fields in PMN are also randomly multivalued. 

It should also be recalled that the displacements in real relaxor systems are not one-dimensional, but are 3-vector. It was realised long ago that vector spin versions of the SK model in a uniform field would exhibit a spin glass transition in a transverse direction as the temperature is lowered \cite{GT}, but with only weak non-ergodicity in the longitudinal direction until a lower crossover temperature \cite{Cragg}. It seems probable that the first of these transition temperatures will persist even for short-range interactions. It has also been observed experimetally \cite{Campbell}. It is also of probable relevance that the effective dipolar interaction in displacive systems is not anisotropic as in the Ising cases of Refs. {\cite{Andresen2013}} and \cite{Andresen2017}}, in which the dipoles are constrained to lie in the z-direction. Rather it has the more general isotropic form \cite{Zhong}; 
$[{\bf{u}}_{i} . {\bf{u}}_{j}- 3 (\hat{{\bf{R}}}_{ij}.{\bf{u}}_{i})(\hat{{{\bf{R}}}}_{ij}.{\bf{u}}_{j})]/{|{{\bf{R}}_{ij}|^3}} .$

\begin{figure*}
\includegraphics*
[width=.420\textwidth,height=.300\textwidth]
{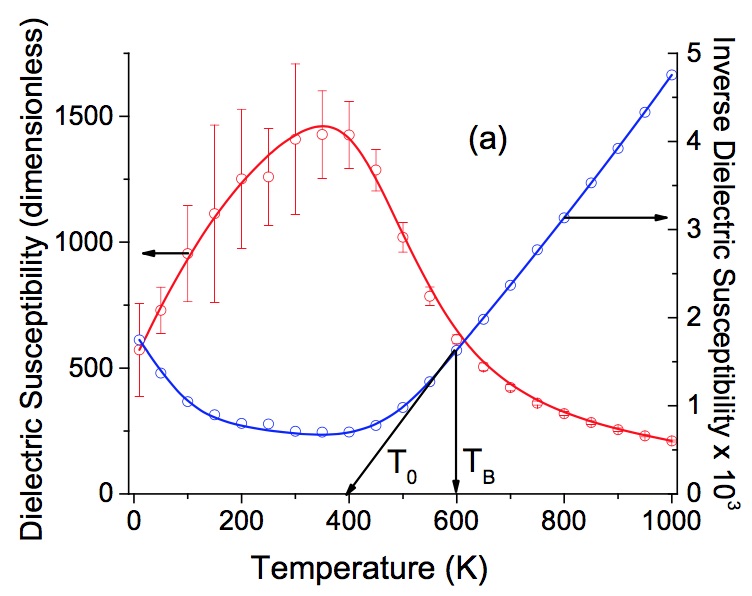}\hfill
\includegraphics*
[width=.420\textwidth,height=.300\textwidth]
{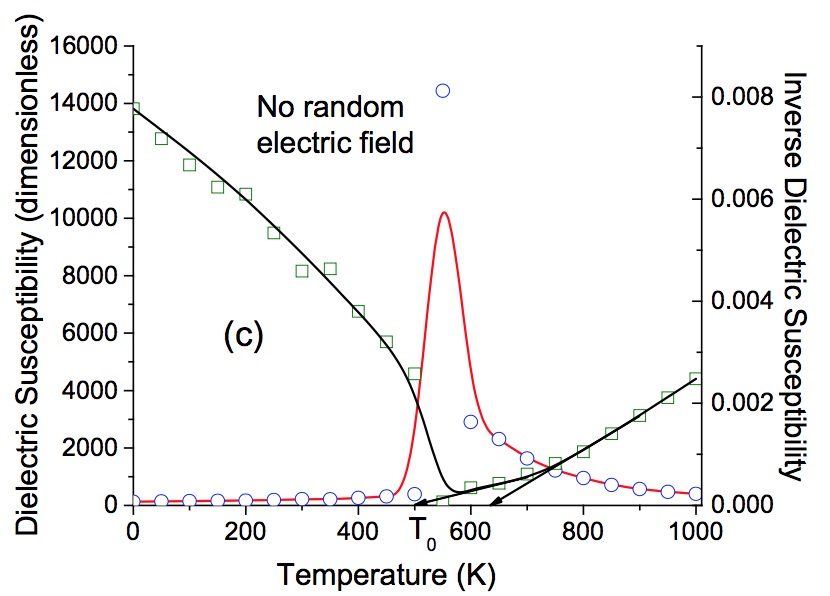}
\caption{
Susceptibilities of model of PMN based on averaged interactions, (a) with random fields, (b) without random fields
\cite{Bellaiche2} \copyright{APS (2015)}.}
\label{fig:Bellaiche_PMN}
\end{figure*}

As already noted, others have claimed that the relaxor peak observed in PMN is driven dominantly by the random fields \cite{Westfal}\cite{Kleemann2}\footnote{See also \cite{Kleemann_this_book}}. A recent simulational study inspired by PMN  has also indicated in favour of this \cite{Bellaiche2}, using a model similar to Eqn. (\ref{eq:PMN1}) with the only disorder attributed to the random field terms; {\it{i.e.}} with $H^{1}_{PM^{*}N^{*}} $ calculated with  parameters averaged over the M and N and ignoring the extra site-interaction terms. Fig {\ref{fig:Bellaiche_PMN}} shows the results obtained for the susceptibility both with and without inclusion of the random fields. Taken
in combination these results are suggestive that relaxor/glass-like behaviour might be possible in the combination of frustrated interactions and disorder of either dilution or random fields. However, as yet, there is no convincing finite-size scaling demonstration of a true thermodynamic transition in the case of purely random field disorder, even when the non-disordered system has a ferro transition. There remains also uncertainty in the statistical mechanics community as to whether there can be a frozen spin glass phase driven purely by random fields without exchange frustration, although it has been proven not to be a thermodynamically stable state in a system of one-dimensional spins with only ferromagnetic (or zero) exchange interactions \cite{Krzakala}.

\subsection{Polar nanoregions}
\label{subsec:PNR}

Another observed feature of displacive relaxors is that of the appearance of polar nanoregions (PNRs)\cite{Cross} \cite{Egami} already at temperatures higher than those of the susceptibility peaks, beneath a higher so-called `Burns temperature' characterised by the onset of deviations from Curie behaviour \cite{Burns}. A commonly expressed conceptualization is that relaxor behaviour is a consequence of interaction of such PNRs, but specific details are not clarified. Here we indicate how such PNRs and  both ferroelectric and relaxor phase transitions can be expected as a consequence of a simple extension of  the modelling above. The initial discussion will be restricted to a simple mean field  consideration and, for simplicity, within a picture allowing only for one-dimensional deviations, but allowing for spatial inhomogeneity. 

We start with the homovalent case. Thus we consider minimization of a Landau-type free energy 
\begin{equation}
F_{R}=\sum_{i} 
\{\tilde{\kappa}_{i}(T) u_{i} ^2
+\tilde{\lambda}_{i}(T)u_{i}^4    
\}
- {    \sum_{(ij)}  \tilde{J}    (R_{ij}, T)   u_{i}u_{j}     }.
\label{eq:LandauF}
\end{equation}
where the coefficients are now temperature-dependent, the $\{u\}$ at minimum are now the mean-field values and allowance is made for different $u_{i}$ at different sites $i$. 

Minimizing with respect to the $\{u_{i}\}$ yields  the self-consistency relation
\begin{equation} 
\tilde{\kappa}_{i}(T) u_{i} 
- {\sum_{j}\tilde{J}(R_{ij})u_{j}} = -2\tilde{\lambda}_{i}u_{i}^3  .
\label{eq:Min F(u)}
\end{equation}
Of particular interest are non-zero solutions and phase transitions as a consequence of reducing the
$\tilde{\kappa}(T)$ with reducing $T$.  
This equation (\ref{eq:Min F(u)}) always allows solutions $\{{u=0}\}$, corresponding to undisplaced paraelectricity, but interest is in possible solutions $\{{u \neq 0}\}$. These only occur for small enough $\tilde{\kappa}$. 

For a pure ferroelectric all the $u_{i}$ have the same value, given by 
\begin{equation}
u=\{
[\sum_{j}{\tilde{J}}(R_{ij},T)-\tilde{\kappa}(T)]/
\tilde{\lambda}(T)
\}^{1/2},
\label{eq:ferroelectric}
\end{equation}
 from which we see that there is a critical temperature $T_{c}$ 
 given by
\begin{equation}
\tilde{\kappa}(T_{c}) =\sum_{j}{\tilde{J}}(R_{ij},T_{c}).
 \label{eq:Tc}
 \end{equation} 
 For
 $T<T_{c}$  the system is ferroelectric
 wheras for  $T > T_{c}$ it is paraelectric.
 
\begin{figure}
\sidecaption
\includegraphics*[width=.600\textwidth,height=.35\textwidth]{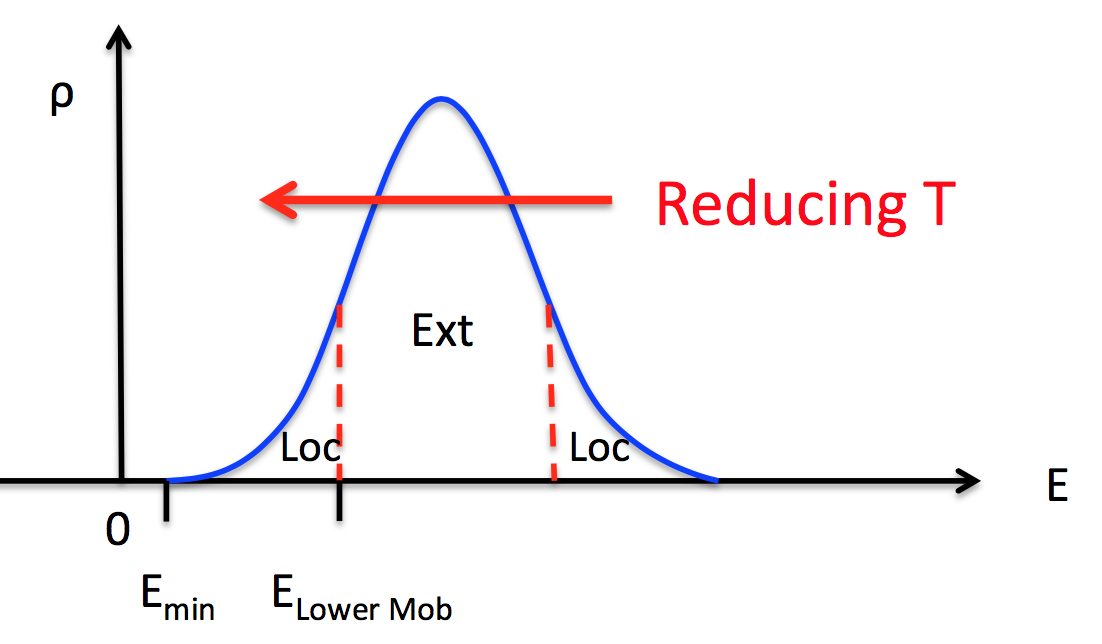}
\caption{Schematic density of states of an Anderson model with local energy disorder, showing localized and extended regions. The arrow indicates movement of the whole figure relative to its vertical axis on decreasing the mean local energy $\epsilon$ while maintaining its relative distribution. \vspace{0.1 cm}}
\label{Anderson_dos}
\end{figure}

 In a general alloy, however, the solutions $u_{i}$ for different sites $i$ will vary. Eqn.({\ref{eq:Min F(u)}}) must have a (real) solution at each site $i$ 
and, in principle, can be either localized or extended/percolating. Localized solutions would represent internally ordered nanoregions, while the onset of extended solutions would signify a phase transition.
A suggestive conceptual guide to the character of such solutions can be visualized by comparing with
the (linear) Anderson localization  equation \cite{Anderson1958}
\begin{equation}
\epsilon_{i} \psi_{i} 
+\sum_{j}  t_{ij} \psi_{j} =E\psi_{i}.
\label{Anderson_loc}
\end{equation}
 with the identifications 
 \begin{equation}
 \{ \epsilon_{i } \}=  \{ {\tilde{\kappa}}_{i} \} \quad ; \quad
\{ t_{ij} \} = -\{ \tilde{J}_{ij} \}.
\label{identification}
\end{equation}
Fig \ref{Anderson_dos} shows a schematic density of states $\rho(E)$ for the Anderson model
in a situation where the lower band edge is positive.. 
Correspondingly the only solution to Eqn.(\ref{eq:Min F(u)}) is ${u=0}$. However, if the temperature 
is reduced so
the mean $\epsilon$ is decreased sufficiently for the lower band edge to reduce below zero, then solutions of  Eqn.(\ref{eq:Min F(u)}) with  $u\neq 0$ exist.. For a pure system with no $\epsilon$-disorder, all the states of Eqn.(\ref{Anderson_loc}) are extended, with the lower band edge state having the highest symmetry, resulting in a phase transition to a state of similar symmetry for Eqn. (\ref{eq:Tc}).
This corresponds to the  ferroelectric transition as found in BT, whereas in BZ $\tilde{\kappa}^{Zr}$ is never small enough for $\rho(E)$  to reach $E=0$.

\begin{figure}
\sidecaption
\includegraphics[width=.50\textwidth,height=.45\textwidth]{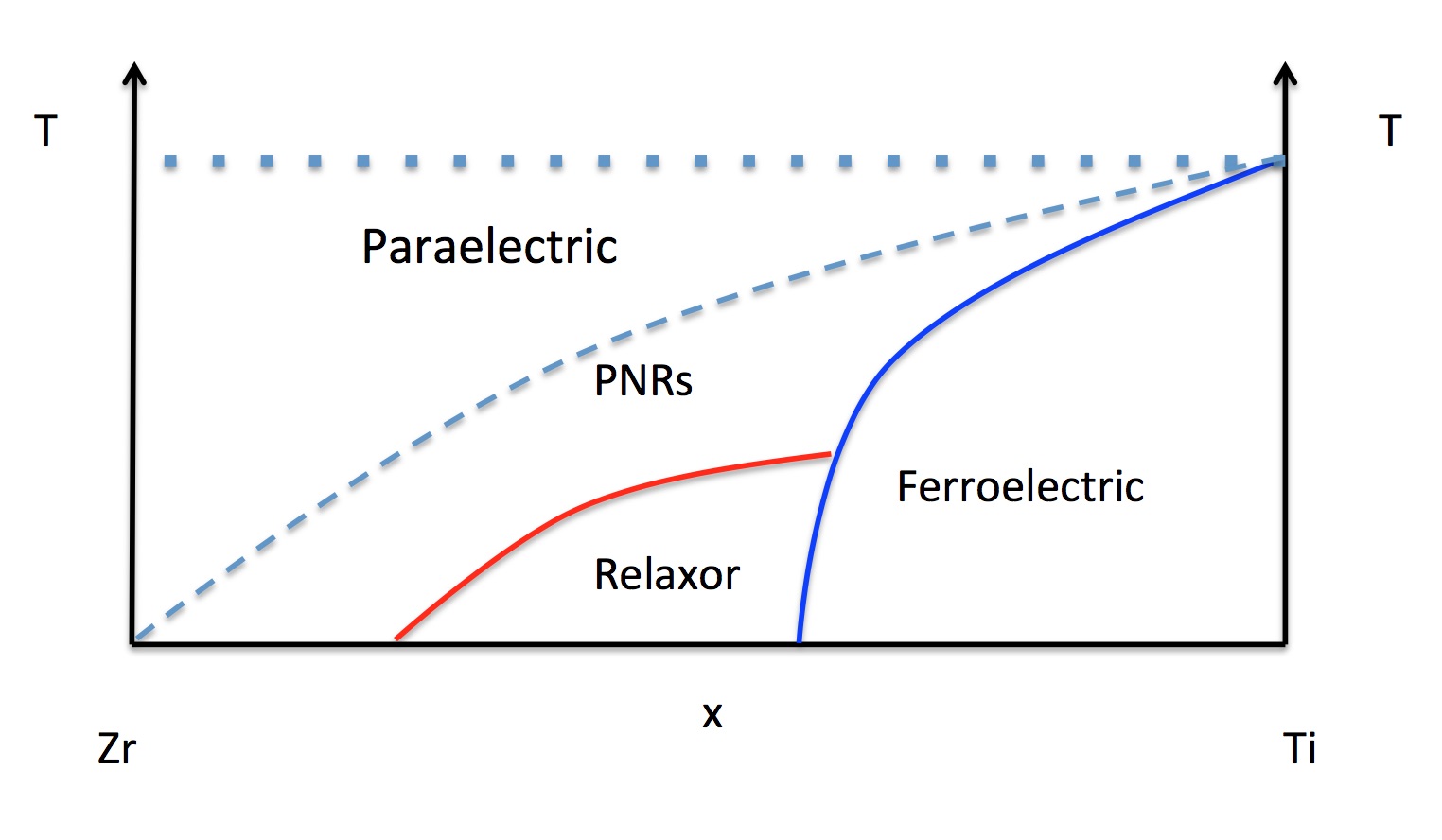}
\caption{Schematic `phase diagram' for BZT expected in the light of heuristic considerations. Solid lines denote true phase transitions. The dotted line indicates the onset of PNRs in the picture discussed. The dashed line is a speculative illustration of crossover for the onset of significant visibility of PRNs. 
\vspace{0.65 cm}}
\label{Schematic_with_PNRs}
\end{figure}

However, if the Anderson model coefficients, such as the $\epsilon$, are disordered then  states at the outer regions of $\rho(E)$ are localized, with the mapping leading to internally-correlated but not cooperatively-frozen clusters, identifiable as the observed PNRs, while for a true thermodynamic phase transition an extended state solution is required. Hence it is (crudely) suggestive that the temperature must be lowered further until the lower mobility edge, separating localized and extended states, crosses $E=0$. This consideration suggests that lowering $T$ in the model system of Eqn.(\ref{eq:Displacive}) will lead first to finite internally ordered nanoregions, growing in number and size as $T$ is lowered, followed by a true thermodynamic transition at a lower temperature. The onset of PNRs is expected at a temperature region close to the phase transition of the pure ferroelectric host. The cooperatively ordering phase transition is expected to be to ferroelectric at higher concentrations of ferroelectric B-ions, passing to relaxor/pseudo-spin-glass at intermediate concentrations, and failing to reach cooperative order at too low concentrations. This is illustrated schematically in Fig. {\ref{Schematic_with_PNRs}}, where the solid lines indicate phase transitions but the dotted and dashed lines are heuristic indications of onset and visibility of PNRs. 
\footnote{Conceptually one can view the situation in a substitutional alloy as follows: (i) quenched statistical fluctuations in the locations of the ions on the underlying lattice will lead to a range of clusterings of the more potentially displacable ions (Ti in BZT), with regions both denser and less dense than the average concentration; (ii) For clusters  to displace-order internally the energy lowering gained through interaction must overcome the local free energy penalties; (iii) such internal correlation  will first occur on clusters that are close in structure to the pure ferroelectric one (BT for BZT);  (iv) this can always occur in principle at a temperature close to that of the pure ferroelectric,  but  will become rarer as the concentration of potentially ferroelectric ions reduces; (v)  as the temperature is lowered the decrease in the effective $\kappa(T)$ will lead to the internal mean-field stabilization of larger clusters, until eventually there will be clusters that percolate throughout the whole system; (vi) the character of the final low temperature macroscopically  cooperative  state will be determined by minimizing the free energy, which in a disordered and frustrated system can be either globally periodic or spin glass-like.}
This prediction including PNRs  is in qualitative accord with experimental observations \cite{Maiti}.

 For heterovalent alloys it is necessary also to take account of the effective random field terms in the Hamiltonian, which yield corresponding  linear contributions to the mean-field free energy of Eqn. (\ref{eq:LandauF}) and consequently terms of zero-order in the $\{u\}$ in Eqn. (\ref{eq:Min F(u)}) and hence  induced displacements even at higher temperatures; {\it{e.g.}} even without any interaction terms the presence of the extra charges on  Mg++ and Nb+++++ ions  randomly-distributed on B-sites of PMN would lead to a corresponding quasi-spherical distribution of Pb deviations from their mean-lattice positions, as discussed in \cite{DS_PMN}, in qualitative accord with observations \cite{Vakhrushev}. Statistical clusterings of effective fields of similar orientations can be expected to lead to nucleation of polar nanoregions, even without frustration in the site-to-site interactions. Although the suggestive quasi-mapping  to the Anderson equation suggested above will no longer be applicable, the concept of relating the `transition' to an `edge' separating localized and extended solutions should remain qualitatively valid.

In principle one could also change description further by considering the PNR as `superspins' with effective interactions betwen them, with eventually a percolating coherence between them marking the relaxor transition. This conceptualization was used in the context of itinerant spin glasses in the early 70's \cite{DS_Mihill1} and has become popular in considerations of relaxors; for a recent discussions see {\it{e.g. }}\cite{Kleemann2} \cite{Kleemann_Stirin} \cite{Kleemann2017} \cite{Kleemann_this_book}.

Of course, to be fully representative of even the soft-Ising-like  model of Eq.(\ref{eq:Displacive}) for a homovalent alloy, one needs to go beyond the simple mean-field form used above.

\section{Itinerant spin glasses}

In fact, some the suggestions above in section {\ref{bsec:3}} were  conceptually pre-empted many decades ago by theoretical considerations of itinerant spin glasses, in which the magnetism resides with conduction electrons \cite{DS_Mihill1} \cite{DS_Mihill2} \cite{Hertz_Stoner}. 
These studies were not pursued but are probably worthy of resurrection here in the  light of \cite{DS_BZT}. Again, we  follow the philosophy of using a  simple  model and approximations for illustration. 

A simple 
Hubbard model for a transition metal alloy is given by the Hamiltonian
\begin{equation}
H_{HA}=\sum_{ij;s=\uparrow,\downarrow} t_{ij} a_{is}^\dagger a_{js} +\sum_{i;s=\uparrow,\downarrow}V_{i} a_{is}^{\dagger}a_{is} + \sum_{i}U_{i}\hat{n}_{i\uparrow}\hat{n}_{i\downarrow}
\label{eq:Hubbard}
\end{equation}
where the $a,a^{\dagger}$ are site-labelled d-electron annihilation and creation operators, $\hat{n}_{is}=a^{\dagger}_{is}a_{is}$ and, in general, the $t_{ij}$, $V_{i}$ and $U_{i}$ depend upon 
the type of atoms at sites {\it{i,j}}. We are concerned with cases in which the electron density is such that the conduction band is only partially filled and the alloys are metallic. 

This can be transformed into a form analogous to that of Eqn. (\ref{eq:Displacive}) with the variables local magnetization and charge  fluctuations. 

We first re-write $\hat{n}_{i\uparrow}\hat{n}_{i\downarrow}$ in terms of complete squares using the identity
\begin{equation}
\hat{n}_{i\uparrow}\hat{n}_{i\downarrow} ={\frac{1}{4}}     \{ \hat{n}_{i}^2 -\hat{m}_{i}^2 \}
\label{eq: transform_to_squares}
\end{equation}
where 
\begin{equation}
\hat{n}_{i}=\hat{n}_{i\uparrow} + \hat{n}_{i\downarrow}; \hspace{0.5 cm}
\hat{m}_{i}=\hat{n}_{i\uparrow} - \hat{n}_{i\downarrow}.
\label{eq:n_and_m}
\end{equation}
For easier conceptualization of the possible magnetic consequences, with minimal more peripheral  distractions,
we further simplify by assuming that the charge fluctuations are of lesser importance   and take their contribution to be absorbed into the  $V_{i}$ and furthermore set all these $V_{i}$ equal and hence ignorable.  
Further re-writing in a symmetric notation, we are left with
\begin{equation}
H=\sum_{ij,\sigma}t_{ij}a_{i\sigma}^\dagger a_{j\sigma} -{\frac{1}{4}}\sum_{i}U_{i}{\hat{\textit{\textbf{S}}}}_{i}.{\hat{\textit{\textbf{S}}}}_{i}
\label{Hubbard_S}
\end{equation}
where
\begin{equation}
{\hat{\textit{\textbf{S}}}}_{i}
=a_{is}^{\dagger}
{\boldsymbol{\sigma}}_{s,s'}a_{is'}.
\end{equation}

The
quadratic form of the 
$\hat{\textit{\textbf{S}}}$-term
 in Eqn.({\ref{Hubbard_S}})
enables the use of an `inverse completion of a square' procedure
\cite{Edwards55} \cite{Gelfand} \cite{Stratonovich} \cite{Hubbard}
\cite{Sherrington2} \cite{Sherrington3}
 to effectvely `linearize' the Hamiltonian in 
$ a_{is}^\dagger a_{js'}$ 
  through the introduction of an auxiliary magnetization field variable $\hat{\textbf{\textit{m}}}$, conjugate to $\hat{\textit{\textbf{S}}}$.

 One can then further `integrate out'  the original electron operators in favour a description in terms purely of magnetization variables \cite{Sherrington3}.  Further taking the static approximation yields  
an effective Hamiltonian in local magnetisation variables \footnote{Note that here the $\hat{\textbf{\textit{m}}}$ are auxiliary field variables, not the actual equilibrium magnetizations.}; to fourth order,
\begin{eqnarray}
&H_{m}= \sum_{i}  (1-U_{i}\chi_{ii})   
|{\hat{\textit{\textbf{m}}}}_{i}|^2 - 
 \sum_{ij;i\neq j}U_{i}^{1/2}U_{j}^{1/2}\chi_{ij}{\hat{\textit{\textbf{m}}}}_{i}.{\hat{\textit{\textbf{m}}}}_{j} \nonumber \\
 &- 
\displaystyle{\sum_{ijkl;\alpha\beta\gamma\delta}}(U_{i}U_{j}U_{k}U_{l})^{1/2}\Pi^{\alpha\beta\gamma\delta}_{ijkl}
\hat{m}^{\alpha}_{i}\hat{m}^{\beta}_{j}\hat{m}^{\gamma}_{k}\hat{m}^{\delta}_{l},
\label{H_Sm}
\end{eqnarray}
where $\chi$ is the static band susceptibility function of the bare system (with only the $t$ term), $\Pi$ is a corresponding  bare 4-point function and we have dropped the higher order contributions.

A further change of variables
\begin{equation}
{\hat{\textit{\textbf{M}}}}_{i}=U_{i}{\hat{\textit{\textbf{m}}}}_{i}
\end{equation}
immediately brings this to a form reminiscent of Eqn.({\ref{eq:Displacive}}):
\begin{eqnarray}
&H_{M}= \sum_{i}  (U_{i}^{-1}-\chi_{ii})   
|{\hat{\textit{\textbf{M}}}}_{i}|^2 - 
 \sum_{ij;i\neq j}
 \chi_{ij}{\hat{\textit{\textbf{M}}}}_{i}.{\hat{\textit{\textbf{M}}}}_{j} \nonumber \\
 &- 
\displaystyle{\sum_{ijkl;\alpha\beta\gamma\delta}}
\Pi^{\alpha\beta\gamma\delta}_{ijkl}
\hat{M}^{\alpha}_{i}\hat{M}^{\beta}_{j}\hat{M}^{\gamma}_{k}\hat{M}^{\delta}_{l},
\label{H_SM}
\end{eqnarray}
with local self-energy  weight $(U_{i}^{-1}-\chi_{ii})$ the analogue of  $\kappa$ in Eqn.({\ref{eq:Displacive}}). Minimization of $H_{m}$ or $H_{M}$ gives the equation for the   the magnetizations \{$\textit{\textbf{m}}$\} in mean field approximation, the itinerant magnetic analogue of the relaxor  Eqn.(\ref{eq:Min F(u)}).

A simple consideration of a system with two components A and B with $U_{A}=0$ but $U_{B }> 0$ immediately demonstrates the following well-known mean field results: (i) 
pure A is only paramagnetic; (ii) pure B is ferromagnetic only if $({1-U_{B}\sum_{j}\chi_{ij})} \equiv ({1-U_{B}\chi(q=0)})<0$, the Stoner criterion \cite{Stoner}, otherwise paramagnetic, (iii)
a single B substituted  in an A-host will only carry a mean-field moment if $({1-U_{B}\chi_{ii})} \equiv ({1-U_{B}\int_{q}\chi(q))}<0$, the Anderson condition \cite{Anderson_1961}.

 For metallic systems $\chi_{ij}$ oscillates in sign as a function of separation, so  there is frustration in the effective interactions of Eqn. ({\ref{H_SM}}). Hence,  a more concentrated alloy with a sufficient  finite non-zero density of $B$ atoms can, in principle, exhibit either ferromagnetism or another periodic order, while. beneath a critical concentration $x_c$ and with sufficient frustration, it can exhibit spin-glass order,
 
 If the Anderson local moment  criterion is  satisfied at B-sites, then the situation is essentially the same as in the conventional hard spin case discussed in Section {\ref{sec:3}.
 
 However, if the Anderson criterion is not satisfied (equivalent to $\kappa > 0$) then a sufficiently strong potential energy lowering due to coherently-acting  mutual magnetization fluctuation freezing at different sites is needed to bootstrap a macroscopic magnetically ordered phase, overcoming the $(1-U^{B}\chi_{ii})|{\bf{m}}_{i}|^2$ local fluctuation penalties on a percolating network, otherwise the system would be paramagnetic.  
 For the case of a high concentation of $B$ this phase is still essentially Stoner's itinerant ferromagnetism. But for an intermediate concentration of $B$ the spontaneous cooperative phase can be a spin glass, bounded by a lower critical concentration separating it from the (Pauli-type) paramagnet and an upper critical concentration separating it from the ferromagnet. Hertz \cite{Hertz_Stoner} provided the first theory\footnote{Using a different formulation than presented here.} and introduced the term `Stoner glass' to refer to the itinerant spin glass.

Furthermore, the same considerations concerning the formation of PNRs as discussed for  displacive ferroelectric alloys should apply to the formation of  bootstrapped super-spin nano-clusters due to quenched statistical fluctuations in the locations of the B atoms, in such itinerant magnetic alloys, even above a transition temperature for spin glass or ferromagnet \cite{DS_Mihill1} \cite{DS_Mihill2}\footnote{These early papers discussed the formation of clusters and anticipated that their interactions would then yield the spin glass state, much as in later considerations of relaxors.}.
Their   `visibility' would however depend on their effective dynamical lifetimes, not discussed here but surely much shorter than those for ferroelectric PNRs.

The sequence paramagnet/spin glass/ferromagnet was already observed in the early days of experimental spin glass physics; {\it{e.g.}} in {\bf{Rh}}Co alloys \cite{Coles-Tari}. It seems highly probable that a similar phenomenological explanation should apply in other metallic spin glass alloys, especially those labelled as `cluster glasses', and 
it could prove interesting to review them in this light. A full theoretical treatment would require going beyond the simple mean-field theory presented above, as well as beyond the other simplifying assumptions employed above, but hopefully it could already  provide a useful starting perspective complementary to those currently employed.

\section{Strain glass}

In this section we consider another analogue of spin glasses,  of  glassy strain distortions in martensitic alloys \cite{Ren} \cite{Sherrington_martensite}  \cite{Sherrington_Glassy}  \cite{Ren2018} and given the name `strain glass'.

Martensitic materials, see e.g. \cite{Bhattacharya} 
\cite{Otsuka}, exhibit first-order structural transitions  from higher to lower symmetry phases as temperature is lowered. 
An example is from high temperature cubic austenite to a lower temperature phase of alternating twin planes of complementary tetragonal character, epitomized by  TiNi which in its pure higher-temperature state is an ordered compound
 of rocksalt structure. Our interest here will be particularly in when this compound is atomically  disordered, for example by randomly altering the balance of Ti and Ni or by replacing some of these atoms by another element ({\it{e.g.}}Fe).
 
Although, in principle, it could be modelled microscopically in terms of atomic displacements, as was done for relaxor problem of section {\ref{bsec:3}}, or  phenomenologicaly at a Landau-Ginsburg level in terms of continuous-valued deviatoric strains, here we shall simply employ a crude discrete pseudo-spin mean-field modelling for illustration. Furthermore, again for simplicity, we shall consider a two-dimensional version in which the local cell structure is describable in terms of a variable $S_{i}$ which takes the values 0, +/-1 corresponding respectively to a  square and two orthogonal rectangular stuctures. The local part of an effective free energy is then given by 
\begin{equation}
F_{L}= \sum_{i}D_{i} S_{i}^2
\end{equation}
where the \{i\} label cells, so that if $D_{i}>0$ then minimizing $F_{L}$ yields $S_{i}=0$, while if $D_{i}<0$ it yields degeneracy $S_{i}=\pm 1$. There are also intercell interaction terms 
\begin{equation}
F_{int}=-\sum_{ij}S_{i}V_{ij}S_{j}
\end{equation}
arising from both short-range neighbouring similarity effects and from longer range St. Venant's compatibility constraints that give (in 2 dimensions) \cite{Lookman}
\begin{equation}
V_{ij}^{StV} \propto 
-\cos{(   4 
\theta{( 
 \textit{  
\textbf{R}}_{ij}  
    )}       
      )}
 / |{\textit{\textbf{R}}}_{ij}|^2
\end{equation}
where $\theta({\textbf{R}})$ is the angle subtended by $\textbf{R}$  at a Cartesian axis of the cubic lattice.

Temperature is emulated by taking the $D$ to be temperature dependent, reducing with reducing temperature. Without the interaction term the $S$-values at the minimum of the free energy will change from $S=0$ to $S=\pm1$ when their corresponding $D$ change from positive to negative. When the interaction is included cooperative bootstrapping  will result in $S=\pm 1$  at sites where the resultant free energy minimization can overcome the local $D$, with the favoured state given by the relative signs of the $\{S_{i}\}$ that yield the lowest free energy. For a pure system one gets the martensite phase of diagonally alternating stripes of $S=+1$ and $S=-1$. For a system with quenched randomness of site occupation one can expect a corresponding quenched randomness of the $\{D_{i}\}$, along with quenched effective random fields.  Clearly this represents a scenario similar to that discussed above for relaxors and spin glasses, namely that for sufficient quenched disorder the frustration,
arising from the antiferroelastic interactions at values of $\theta$ where the cosine is negative,
  leads to an expectation  of strain glass, as has been observed  \cite{Saxena} \cite{Ren_2}; see Fig {\ref{fig:NiTi}}.\footnote{In this case the analogue of the ferromagnet is the martensitic stripe phase.}
\begin{figure}
\includegraphics[width=.450\textwidth,height=.45\textwidth]{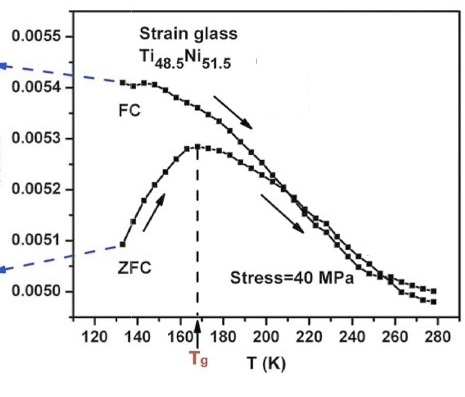}\hfill
\includegraphics[width=.550\textwidth,height=.45\textwidth]{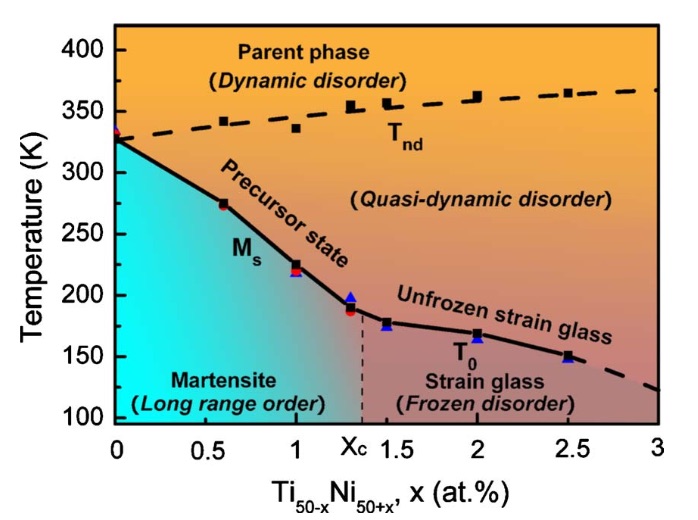}
\caption{${Ni_{50-x}Ti_{50+x}}$:
(a) FC/ZFC evidence for strain glass \cite{Saxena} \copyright{APS (2007)} (b) Phase diagram \cite{Ren_2}\copyright{APS (2010)}}
\label{fig:NiTi}
\end{figure}
Both the cases of quenched $D$ \cite{Sherrington_martensite} and of the effective random fields \cite{Wang}, arising from local substitutions, have been proposed separately as the disorders responsible  strain glass behaviour. The reality probably includes both.

\section{Conclusion}

In this chapter I have tried to demonstrate similarities in the potential for ferroic glass behaviour in several different types of system, magnetic, ferroelectric and martensitic/ferrolastic alloys, both metallic and insulating, using a combination of simple modelling and analogies, experimental, theoretical/mathematical, comutational and conceptual, and with a particular consideration of `induced moment' and continuously displaceable systems. 
   
The key ingredients to permit such glassy behaviour appear to be frustrated interactions and quenched disorder, as has often been expressed before. Probing these relationships has provided possible explanations for phenomena such as the onset of non-ergodicity and slow dynamics. Cluster effects such as those known as polar nano-regions (PNR) in  displacive relaxors are considered in analogy with localization phenomena.

  The  simple analogies considered here suggest further conceptual transfers between different ferroic materials and further experimental investigations; for further discussion concerning characteristic experimental  aspects of spin glasses see {\cite{Vincent}; for a complemetary recent discussion of relaxors see \cite{Kleemann_this_book}.

These comparisons have also highlighted some remaining questions, particularly concerning the issue  of the role of quenched random fields. BZT has no quenched random fields but PMN has significant such fields, yet the susceptibility  measurements look very similar to one another.

    As noted earlier, there is controversy in the spin glass community as to whether a true spin glass phase transition can continue to exist in the presence of an applied field.
Even without a non-analyticity it 
    would not necessarily mean that the peak in the pure zero-field susceptibility cannot continue in a finite-field, in a more rounded form,
     as indeed  was clear already in the early important  small-field experiments of Cannella and Mydosh \cite{CannellaMydosh}; see Fig {\ref{Susc_ferro_BZT_AuFe}}(b). There has been much interest in the random field Ising model (RFIM), normally with short-range ferromagnetic interactions, without demonstration of a spin glass, and indeed it has been proven not to be thermodynamically stable for purely ferromagnetic or zero interactions \cite{Krzakala}. There have been theoretical suggestions that in a system with higher spin dimension there could be a `spin glass' state driven by the random fields \cite{Mezard-Monasson} but there has been no observed evidence of a sharp transition to such a phase in a magnetic system. 

Although there have been many experimental demonstrations of spin glass behaviour in frustrated and quench-disordered  3-dimensional systems of 3-dimensional (Heisenberg) spins, there is still some debate about theory \cite{Kawamura}. There are no simple experimental methods to apply three-dimensional random magnetic fields On the other hand,  the relaxor systems discussed above have displacement variables able to orient in the full 3-dimensional space and in heterovalent relaxor alloys, such as PMN,  the effective random fields  are also spread throughout the 3-dimensional orientation space and are of significant strength, yet the peaks in the susceptibility  are quite sharp. Both the classic spin glass and the relaxor examples have long range interaction frustration.

It is thus tempting to wonder whether the criterion of frustrated interaction and quenched disorder as the key ingredients for spin glass/ relaxor/ strain glass behaviour might apply independently whether the disorder arises from site-disorder, bond-disorder or random fields, or a combination, preventing simple homogenous and smoothly varying optimal compromises, and also whether one needs to go beyond one-dimensionality of the  `pseudospins', but  more work is required to help decide.

Finally, let me note that my aim has not been to describe quantitavely or completely the systems that I have discussed, but rather. through simple extraction and comparisons, to try to draw links and to expose contrasts and remaining puzzles and uncertainties, in the hope that  they might stimulate work that might not have been obvious within the confines of just sub-classes of systems. I  should also point out that I am not the first to propose that either relaxors or martensitic alloys might be considered as pseudo-spin glasses (see {\it{e.g.}}\cite{Viehland} \cite{Blinc-Pirc} \cite{Sethna}), but I hope my small contribution can be stimulating in moving towards a greater understanding.


\begin{thebibliography}{12}

\bibitem{Weiss} 
Weiss P,  
 L'hypoth\`ese du champ mol\'eculaire et la propri\'et\'e ferromagn\'etique, 
J. Phys. Theor. Appl.{ \bf{6}}, 661 (19

\bibitem{Stoner}
Stoner EC,  
Collective electron ferromagnetism,
 Proc Roy. Soc. A,{ \bf{165}}, 372 (1938)

\bibitem{Valasek}
Valasek J, 
Piezoelectric and allied phenomena in Rochelle salt,
 Phys. Rev {\bf{15}}, 537 (1920)

\bibitem{Neel2}
N\'eel L,
 Propri\'et\'ees magn\'etiques des ferrites; Ferrimagn\'etisme et antiferromagn\'etisme, 
Ann. Phys. Paris {\bf{3}},  137 (1948)

\bibitem{CannellaMydosh}
Cannella V and Mydosh J A,
Magnetic Ordering in Gold-Iron Alloys.
  Phys.Rev.B {\bf{6}}, 4220 (1972)


 
 \bibitem{EA}%
Edwards SF and   Anderson PW, 
Theory of spin glasses,
 J.Phys.F. \textbf{5}, 965 (1975).

\bibitem{SK}%
Sherrington D and   Kirkpatrick S,
 Solvable Model of a Spin-Glass,
 Phys.Rev.Lett. \textbf{35}, 1792 (1975).
 
 \bibitem{Parisi}
 Parisi G,
Infinite Number of Order Parameters for Spin-Glasses,
  Phys. Rev. Lett. {\bf{43}}, 1754 (1979)
 
  \bibitem{Mezard}%
 M\'ezard M, Parisi G  and Virasoro M-A,
{\it{Spin Glass Theory and Beyond}} (Word Scientific, Singapore, 1987)
 
\bibitem{Binder-Young}%
Binder K  and Young AP,
 Spin glasses: Experimental facts, theoretical concepts and open questions, 
Rev. Mod. Phys. \textbf{58}, 801 (1986)
 

\bibitem{Fischer-Hertz}%
Fischer KH and Hertz JA, 
{\it{Spin Glasses}} (Cambridge University Press, Cambridge 1991)

\bibitem{Mydosh1}%
Mydosh JA, {\it{Spin Glasses: An Experimental Introduction}} (Taylor and Francis, Philadephia 1993)

\bibitem{Nishimori}
Nishimori H, {\it{Statistical Physics of Spin Glasses and Information Processing}} (Oxford University Press, New York 2001)
 
 \bibitem{Mydosh2}
Mydosh JA, 
Spin glasses: redux: an updated experimental/materials survey,
 Rep. Prog. Phys. {\textbf{78}},  052501 (2015)

\bibitem{Kawamura} 
Kawamura H and Taniguchi T, 
Spin Glasses, in {\it{Handbook of Magnetic Materials}}, ed. Buschov KHJ (Elsevier BV, Amsterdam 2015)

\bibitem{Panchenko}
Panchenko D, { \it{The Sherrington-Kirkpatrick Model} }(Springer, New York 2013)

\bibitem{Smolenskii}%
Smolenskii GA and Isupov VA, 
Dokl. Acad. Nauk SSSR, {\textbf{97}}, 653 (1954)

\bibitem{Smolenskii2}%
Smolenskii, GA, Isupov VA,  Agranovskaya AI and Popov SN,
Ferroelectrics with diffuse phase transitions, Fiz.Tverd Tela  {\textbf{2}}, 2906 (1960);~~[Sov.Phys.Solid State \textbf{2}, 2584 (1961)]

\bibitem {Ren}
Sarkar S,  Ren X, and  Otsuka K., 
Evidence for Strain Glass in the Ferroelastic-Martensitic System $\mathrm {Ti}_{50-x}\mathrm{NI}_{50+x}$, 
Phys. Rev. Lett. {\bf{95}}, 205702 (2005)

\bibitem{Wassermann}%
 Lecomte GV, von L{\"o}hneysen H and Wassermann EF, 
Frequency Dependent Magnetic Susceptibility and Spin Glass Freezingin {\underline{Pt}}Mn Alloys,
{Z.Phys. B} {\textbf{50}}, 239 (1983)
 


\bibitem{Maglione}
Simon A, Ravez J and Maglione M, 
The crossover from a ferroelectric to a relaxor state in lead-free solid solutions,
J.Phys.Cond.Mat. {\bf{16}}, 963 (2004)



\bibitem{Levstik}%
 Levstik A, Kutnjak Z, Filipi\v{c} C and Pirc R, 
 Glassy freezing in relaxor ferroelectric lead magnesium niobate,
 {Phys.Rev.B} {\textbf{57}}, 11204 (1998)
 
 \bibitem{Nagata}%
Nagata S, Keesom PH and Harrison HR, 
Low-dc-field susceptibility of {\it{Cu}}Mn spin glass,
Phys.Rev. B \textbf{19}, 1633 (1979)
 
\bibitem{Akbarzadeh2012}%
Akbarzadeh AR, Prosandeev S, Walter EJ, Al-Barakaty A and Bellaiche L, 
Finite-Temperature Properties of ${\mathrm{Ba(Zr, Ti) O}_3}$ Relaxors from First Principles,
{Phys.Rev.Lett.} {\textbf {108}}, 257601 (2012)


\bibitem{Coles_AuFe}%
The role of finite magnetic clusters
in Au-Fe alloys near the percolation
concentration,
Coles BR, Sarkissian B and Taylor RH, Phil.Mag.B {\textbf{37}}, 489 (1978)

\bibitem{Maletta}%
Maletta H and Convert P,
Onset of Ferromagnetism in $Eu_{x}Sr_{1-x}S$ near x=0.5,
 Phys.Rev.Lett. \textbf{42}, 108 (1979)


\bibitem{Sherrington_Glassy}%
Sherrington D., Understanding glassy phenomena in materials
 in {\it{Disorder and Strain induced complexity in functional materials}}, eds.
Kakeshita T., Fukuda T., Saxena A. and Planes A.,
177 (Springer, Berlin 2012)

\bibitem{Toulouse}
Toulouse G, 
Theory of frustration effect in spin-glasses, 
Comm.Phys {\textbf{2}}, 115 (1977)


\bibitem{SS}
Sherrington D and Southern BW,
Spin glass versus ferromagnet,
  J.Phs.F, {\bf{5}}, L49 (1975)

\bibitem{Binder}
Binder K., 
 Finite Size Scaling Analysis of Ising Model Block Distribution Functions, 
 Z.Phys.B {\bf{43}}, 119 (1981)

\bibitem{Bhatt-Young}
Bhatt RN and Young AP, 
Search for a transition in the 3-dimensional $\pm 1$ spin-glass
Phys.Rev.Lett. {\bf{54}} 924 (1985)

\bibitem{Katzgraber-Young}
Katzgraber HG, Koerner M, and Young AP, 
Universality in three-dimensional Ising spin glasses: A Monte Carlo study,
Phys.Rev B {\bf{73}} 224432 (2006)

\bibitem{Katzgraber_PRX}
Andresen JC, Katzgraber HG, Organesyan V and Schechter M, 
Existence of a Thermodynamic Spin-Glass Phase in the Zero-Concentration Limit
of Anisotropic Dipolar Systems,
Phys.Rev.X {\bf{4}}, 041016 (2014)


\bibitem{Sarkissian-Coles}
Sarkissian BVB and Coles BR, 
Spin-glass to Overhauser-alloy transitions in Y-rare-earth and Sc-rare-earth  alloys,
Comm.Phys.1, 17 (1976)


\bibitem{Ghatak}
Ghatak SK and Sherrington D,
Crystal field effects in a general S Ising spin glass
  J.Phys.C {\bf{10}}, 3149 (1977)

\bibitem{Hochli}
Hochli UT, Knorr K and Loidl A, 
Orientational glasses,
Adv.Phys. {\bf{1990}}, 405 (1990)

\bibitem{Vugmeister}
Vugmeister BE and Glinchuk MD,
Dipole glass and ferroelectricity in random-site electric dipole systems,
Revs.Mod.Phys. {\bf{62}},993 (1990)

\bibitem{Goldbart}
Goldbart PM and Sherrington D, 
Replica theory of the uniaxial quadrupolar glass
J.Phys.C {\bf{18}}, 1923 (1985)

\bibitem{Sherrington-Japan}
Sherrington D, 
Potts and Related Glasses,
Prog.Theor.Phys.Japan Supp.{\bf}{87}, 180 (1986)


  


\bibitem{Binder-Reger}
Binder K and Reger JD,
Theory of orientational glasses models, concepts, simulations,
  Adv.Phys. {\bf{41}}, 547 (1992)






\bibitem{Zhong}
Zhong W, Vanderbilt D and Rabe KM,
Phase Transitions in $\mathrm{BaTiO}_3$ from First Principles,
Phys. Rev. Lett. 73, 1861 (1994);
First-principles theory of ferroelectric phase transitions for perovskites: The case of $\mathrm{BaTiO}_3$,
Phys.Rev.B {\bf{52}}, 6301 (1995)


\bibitem{King-Smith}
King-Smith RD and Vanderbilt D,
First-principles investigation of ferroelectricity in perovskite compounds,
Phys.Rev.B {\bf{49}}, 5828 (1994)


\bibitem{Maiti}%
Maiti T, Guo R and Bhalla AS, 
Structure-Property Phase Diagram of $\mathrm{BaZr_{x}Ti_{1-x}}O_{3}$ System
{J.Am.Cer.Soc.} {\textbf 91}, 1769 (2008)



\bibitem{DS_BZT} 
Sherrington D, 
BZT: A Soft Pseudospin Glass,
Phys.Rev.Lett, {\bf{111}}, 227601 (2013)

 \bibitem{Kleemann-Miga}%
 Kleemann W, Miga S, Dec J and Zai J,
Crossover from ferroelectric to relaxor and cluster glass in $\mathrm{BaTi}_{(1- x)}\mathrm{Zr}_{x}\mathrm{O}_3 (x=0.25 - 0.35)$ studied by non-linear permittivity,
{App.Phys.Lett.} {\textbf{102}}, 232907 (2013)

\bibitem{Luttinger_Tisza}
Luttinger JM and Tisza L, 
Theory of Dipole Interaction in Crystals,
Phys.Rev. {\bf{70}}, 954 (1946)

\bibitem{Fernandez}
Alonso JJ and Fernandez JF,
Monte Carlo study of the spin-glass phase of the site-diluted dipolar Ising model,
  Phys.Rev.B {\bf{81}}, 064408 (2010)
  



\bibitem{Andresen_PRX}
Andresen JC, Katzgraber HG, Organesyan V and Schechter M,
Existence of a Thermodynamic Spin-Glass Phase in the Zero-Concentration Limit
of Anisotropic Dipolar Systems,
  Phys.Rev X  {\bf{4}}, 041016 (2014)
  
  \bibitem{Hewat}
 Hewat AW,
 Structure of rhombohedral ferroelectric barium titanate,
  Ferroelectrics {\bf{6}}, 215 (1974)
  
 \bibitem{Shirane}
 Shirane g, Pepinsky R and Frazer BC, 
 X-ray and neutron diffraction study of ferroelectric $\mathrm{PbTiO_3}$, 
  Acta Crystallogr. {\bf{9}}, 131 (1955)


\bibitem{DS_Stirin}
Sherrington D,
Relaxors, spin, Stoner and cluster glasses,
  Phase Trans.{ \bf{88}}, 202 (2015)
  

  
\bibitem{Westfal}%
Westphal V, Kleemann W  and Glinchuk MD, 
Diffuse Phase Transitions and Random-Field-Induced Domain States of the "Relaxor" Ferroelectric $\mathrm{PbMg_{1/3}Nb_{2/3}0_{3}}$
{Phys.Rev.Lett.} {\textbf{68}}, 847 (1992)

\bibitem{Kleemann2}
Kleemann W,
Relaxor ferroelectrics: Cluster glass ground state via random fields and random bonds,
 Phys.Stat.Sol. B {\bf{251}}, 1993 (2014)
 
 \bibitem{Kleemann_this_book}
Kleeman W,
Relaxor ferroelectrics and related cluster glasses
This book, chapter ??? (2017)


\bibitem{Cowley}%
Cowley RA, Gvasaliya SN, Lushnikov SG, Roessli B and Rotaru GM, 
Relaxing with Relaxors,
Adv.Phys. \textbf{60}, 229 (2011)

\bibitem{Almeida-Thouless}
de Almeida JRL and Thouless DJ, 
Stability of the Sherrington-Kirkpatrick solution of a spin glass model,
J.Phys.A {\bf{11}}, 983 (1978)

\bibitem{Sharma-Young}
Sharma A and Young AP, 
de Almeida–Thouless line in vector spin glasses
Phys.Rev.E {\bf{81}}, 061115  (2010) 

\bibitem{GT} 
Gabay M and Toulouse G, 
Coexistence of Spin-Glass and Ferromagnetic Orderings,
Phys.Rev.Lett. {\bf{47}}, 201 (1981)

\bibitem{Cragg}
Cragg DM, Sherrington D and Gabay M, 
Instabilities of an {\it{m}}-Vector Spin Glass in a Field,
Phys.Rev.Lett. {\bf{49}}, 158 (1982)










\bibitem{Katzgraber-Young2}
Katzgraber HG and Young AP, 
Probing the Almeida-Thouless line away from the mean-field model,
Phys.Rev.B {\bf{72}}, 184416 (2005)

\bibitem{Larson}
Larsen D, Katzgraber HG, Moore MA and Young AP, 
Spin glasses in a field: Three and four dimensions as seen from one space dimension,
Phys.Rev.B {\bf{87}}, 024414 (2013)

\bibitem{Janus_field}
M Baity-Jesi M et al. (JANUS),
The three-dimensional Ising spin glass in an external magnetic field: the role of the silent majority,
  J.Stat.Mech.{ \bf{2014}}, P05014 (2014)


\bibitem{Singh-Young}
Singh RRP and Young AP, 
de Almeida–Thouless instability in short-range Ising spin glasses,
Phys.Rev.E {\bf{96}}, 012127 (2017)




\bibitem{Andresen2013}
Andresen JC, Thomas CK, Katzgraber HG and Schechter M, 
Novel Disordering Mechanism in Ferromagnetic Systems with Competing Interactions,
Phys.Rev.Lett., {\bf{111}}, 177202 (2013)

\bibitem{Andresen2017}
Andresen JC, Katzgraber HG and Schechter M, 
Random-field-induced disordering mechanism in a disordered ferromagnet: Between the Imry-Ma and the standard disordering mechanism,
ArXiv 1706.07904 (2017)

\bibitem{Fernandez_H}
Fernandez JF, 
Evidence against an Almeida-Thouless line in disordered systems of Ising dipoles
Phys.Rev.B {\bf{82}}, 144436 (2010)

    \bibitem{Campbell} 
 Petit D, Fruchter L and Campbell IA,
Ordering in Heisenberg Spin Glasses.
Phys.Rev.Lett. 88, 207206 (2002)

\bibitem{Bellaiche2}
Al-Barakaty A,  Prosandeev, S, Wang D,  Dkhil B and Bellaiche L, 
Finite-temperature properties of the relaxor $PbMg_{1/3}Nb_{2/3}O_{3}$ from atomistic simulations, 
Phys.Rev. B {\bf{91}}, 214117 (2015)

\bibitem{Krzakala}
Krzakala F,  Ricci-Tersenghi F and  Zdeborová L
 Elusive Spin-Glass Phase in the Random Field Ising Model,
Phys. Rev. Lett. 104, 207208


\bibitem{Cross}
Cross LE,
Relaxor ferroelectrics,
  Ferroelectrics {\bf{76}}, 241 (1987)

\bibitem{Egami}%
Jeong IK, Darling TW, Lee JK. 
Proffen T,  Heffner RH,
Park  JS, Hong KS,
Dmowski W and
 Egami T, 
 Local Lattice Dynamics and the Origin of the Relaxor Ferroelectric Behavior,
 Phys.Rev.Lett. \textbf{94}, 147602 (2005)
 
 \bibitem {Burns}
Burns G and  Dacol F, 
Crystalline ferroelectrics with glassy polarization behavior,
    Phys.l Rev. B {\bf{28}} 2527 (1983)

\bibitem{Anderson1958}%
Anderson PW, 
Absence of Diffusion in Certain Random Lattices,
{Phys.Rev.} {\bf{109}}, 1492 (1958)

\bibitem{DS_PMN}
Sherrington D,
$\mathrm {Pb(Mg}_{1/3}\mathrm{Nb}_{2/3}\mathrm{)O}_3$:  a minimal induced-moment soft pseudo-spin glass perspective,
  Phys.Rev.B {\bf{89}}  064105 (2014)


 


\bibitem{Vakhrushev} 
Vakhrushev SB and Okuneva NM, 
Evolution of Structure of $\mathrm{PbMg{1/3}Nb_{2/3}O_3}$ in the Vicinity of the Burns Temperature,
{AIP Conference Proceedings} {\bf{626}}, 117 (2002)



\bibitem{DS_Mihill1}
Sherrington D and Mihill K, 
Effects of clustering on the magnetic properties of transition metal alloys
J.Physique Colloq. {\textbf{35}}, C4-199 (1974)



\bibitem{Kleemann_Stirin}
Kleemann W, Dec J, Miga S,
The cluster glass route of relaxor ferroelectrics,
Phase Trans. {\bf{88}}, 234 (2015)


\bibitem{Kleemann2017}
Kleemann W and Dec J, 
Ferroic superglasses: Polar nanoregions in relaxor ferroelectric PMN 
versus CoFe superspins in a discontinuous multilayer,
Phys. Rev. B  {\bf{94}}, 174203 (2016)





\bibitem{DS_Mihill2}
Sherrington D and Mihill K,
Magnetic ordering in transition metal alloys,
 Proc.Int. Conf.Mag (Moscow 1973) {\bf{1}}, 283 (1974)
 
 \bibitem{Hertz_Stoner}
 Hertz JA, 
 The Stoner glass,
 Phys.Rev. B, {\bf{19}}, 4796 (1979)
 
\bibitem{Edwards55}
Edwards SF, 
The nucleon Green function in pseudoscalar meson theory. I 
 Proc.Roy.Soc. A, {\bf{232}}, 371 (1955)
 
 \bibitem{Gelfand}
Gel'fand IM and Yaglom AM,
Integation in Functional Spaces and its Applications in Quantum Physics,
Uspekhi Mat. Nauk, {\bf{11}},  77 (1956); J.Math.Phys. {\bf{1}}, 48 (1960)


\bibitem{Stratonovich}%
Stratonovich RL,
A method for the computation of quantum distribution functions,
  Doklady Acad.Nauk SSSR {\textbf{115}}, 1097 (1957); Sov.Phys.Doklady {\textbf{2}}, 416 (1958)

\bibitem{Hubbard}%
Hubbard J, 
Calculation of partition functions
Phys.Rev.Lett. {\textbf{3}}, 77 (1959)

\bibitem{Sherrington2}%
Sherrington D,
A new method of expansion in the quantum many-body problem III.The density field,
 Proc.Phys.Soc. {\textbf{91}}, 285 (1967)

\bibitem{Sherrington3}
Sherrington D, 
Auxiliary fields and linear response in Lagrangian many body theory.
J.Phys.C {\bf{4}}, 401 (1971)

\bibitem{Anderson_1961}%
Anderson PW, {Phys.Rev.} {\bf{124}}, 41 (1961)

\bibitem{Coles-Tari}
Coles BR, Tari A and  Jamieson HA, 
Onset of Ferromagnetism in Alloys at Low Temperatures
in {\it{Low-Temperature Physics-LT13}}, ed.  Timmerhaus KD, O' Sullivan WJ and Hammel EF, {\bf{2}} , 414 (Plenum. , New York 1974), %


\bibitem{Sherrington_martensite}
Sherrington D,
A simple spin glass perspective on martensitic shape-memory alloys, 
J. Phys.: Condens. Matter {\bf{20}}, 304213 (2008) 

\bibitem{Ren2018} Ji Y, Ren S, Wang D, Wang Y and Ren, X, Strain glasses, in {\it{Frustrated Materials and Ferroic Glasses, }}
ed. by T. Lookman, X. Ren. Springer Series in Materials Science, vol. 275 (Springer, New York, 2018)

\bibitem{Bhattacharya}%
Bhattacharya K, {\it{ Microstructure of Martensite}} (Oxford University Press, Oxford  2003)

\bibitem{Otsuka}%
Otsuka K and Wayman  CM (eds.), {\it{Shape Memory Materials}} (Cambridge University Press, Cambridge 1998)

\bibitem{Lookman}%
Lookman T, Shenoy SR,  Rasmussen KO, Saxena A and Bishop AR, 
Ferroelastic dynamics and strain compatibility
Phys.Rev.B {\textbf{67}}, 0241114 (2003)

\bibitem{Saxena}
Wang Y,  Ren X,   Otsuka K, and Saxena, A, 
Evidence for broken ergodicity in strain glass,
Phys.Rev.B {\bf{76}}, 132201 (2007) 

\bibitem{Ren_2}%
Zhang Z, Wang Y, Wang D,  Zhou Y,  Otsuka K, and Ren X,
Phase diagram of ${\textrm{Ti}_{50−x}Ni_{50+x}}$: Crossover from martensite to strain glass
Phys.Rev.B {\bf{81}}, 224102 (2010)



 
 \bibitem{Wang}
 Wang D, Wang Y, Zhang Z and Ren X, 
 Modeling Abnormal Strain States in Ferroelastic Systems: The Role of Point Defects,
 Phys.Rev.Lett {\bf{105}}, 205702 (2010)
 
 \bibitem{Vincent}
 Vincent E and Dupuis V,
 Spin glasses: experimental signatures and salient outcomes,
 This volume. Chapter ....... (2017)
 
 \bibitem{Mezard-Monasson}
 M\'{e}zard M and Monasson R,
 Glassy transition in the three-dimensional random-field Ising model,
 Phs.Rev.B {\bf{50}}, 7199 (1994)

\bibitem{Viehland}
Viehland D, Li JF, Jang SJ, Cross LE and Wuttig M.,
Glassy polarization behavior of relaxor ferroelectrics,
Phys.Rev.B {\bf {46}}, 8013 (1992)

\bibitem{Blinc-Pirc}
Pirc R and Blinc R,
Spherical random-bond–random-field model of relaxor ferroelectrics,
Phys.Rev.B {\bf{60}}, 13470 (1999)

\bibitem{Sethna}
Kartha S, Castan T, Krumhansl J A and Sethna J P 
Spin-glass nature of tweed precursors in martensitic transformations
Phys.Rev. Lett. 67 3630 (1991)




\end{thebibliography}
\end{document}